\documentclass[annual]{acmsiggraph}

\usepackage{graphicx}
\usepackage{amsmath}
\usepackage{amssymb}
\usepackage{color}
\usepackage[abs]{overpic}
\usepackage{rotating}
\usepackage{colortbl}
\usepackage{xcolor}
\usepackage[]{algorithm2e}

\definecolor{darkyellow}{rgb}{1,.9,0}

\def\argmin{\mathop{\rm argmin}}

\newcommand{\boldhead}[1]{\vspace{0.05in}\noindent\textbf{#1.}}

\newcommand{\comment}[1]{}
\newcommand{\hili}[1]{{\bf #1}}

\newcommand{\newtext}[1]{{\color{black}{#1}}}
\newcommand{\newnewtext}[1]{{\color{black}{#1}}}

\usepackage{hyphenat}
\TOGonlineid{0131}
\TOGvolume{33}
\TOGnumber{6}
\TOGarticleDOI{2661229.2661256}
\TOGprojectURL{http://kevinkarsch.com/constructaide}
\TOGvideoURL{}
\TOGdataURL{}
\TOGcodeURL{}

\title{ConstructAide: Analyzing and Visualizing Construction Sites\\ through Photographs and Building Models}

\author{
Kevin Karsch \hspace{15mm} Mani Golparvar-Fard \hspace{15mm} David Forsyth
\\ University of Illinois}
\pdfauthor{Kevin Karsch}


\begin{document}

%-------------------
\teaser{
\vspace{-5mm}
\includegraphics[page=1,width=0.105\linewidth,clip=true,trim=350 -50 0 0]{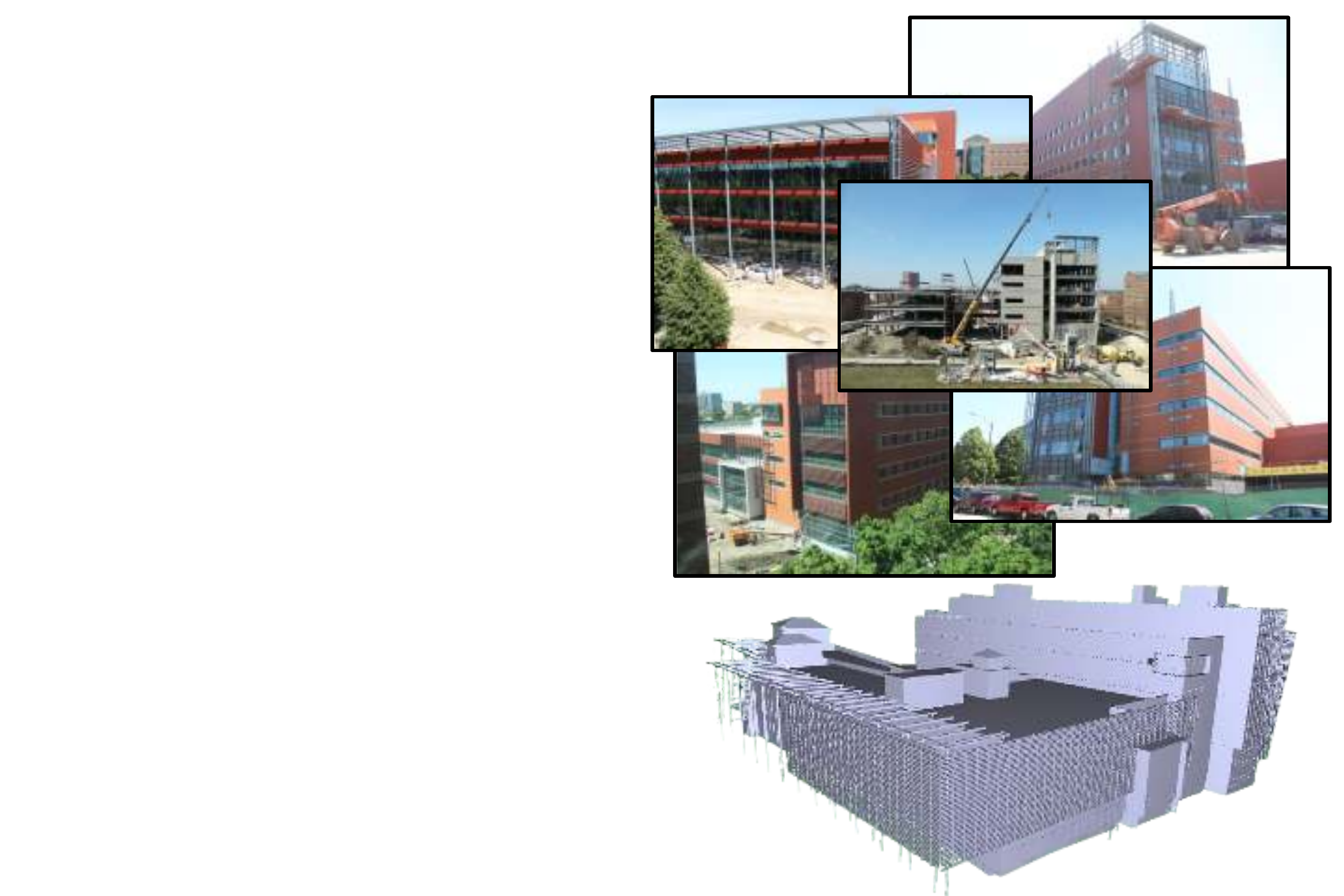}
\includegraphics[page=6,width=0.2175\linewidth,clip=true,trim=0 -50 0 0]{fig/teaser-new.pdf}
\includegraphics[page=5,width=0.2175\linewidth,clip=true,trim=40 0 80 0]{fig/teaser-new.pdf}
\includegraphics[page=3,width=0.2175\linewidth,clip=true,trim=40 0 80 0]{fig/teaser-new.pdf}
\includegraphics[page=4,width=0.2175\linewidth,clip=true,trim=40 0 80 0]{fig/teaser-new.pdf}
\\
\begin{minipage}{0.105\linewidth}\centerline{\large (a)}\end{minipage}
\begin{minipage}{0.2175\linewidth}\centerline{\large (b)}\end{minipage}
\begin{minipage}{0.2175\linewidth}\centerline{\large (c)}\end{minipage}
\begin{minipage}{0.2175\linewidth}\centerline{\large (d)}\end{minipage}
\begin{minipage}{0.2175\linewidth}\centerline{\large (e)}\end{minipage}
\caption{Our system aligns sets of photographs with 4D building models (a) to allow for new modes of construction-site interaction and visualization (using the ConstructAide GUI, b), such as architectural renderings (c), 4D navigation (d), and performance monitoring (e).}
\label{fig:teaser}
}
%-------------------

\maketitle

\begin{abstract}
We describe a set of tools for analyzing, visualizing, and assessing architectural/construction progress with unordered photo collections and 3D building models. With our interface, a user guides the registration of the model in one of the images, and our system automatically computes the alignment for the rest of the photos using a novel Structure-from-Motion (SfM) technique; images with nearby viewpoints are also brought into alignment with each other. After aligning the photo(s) and model(s), our system allows a user, such as a project manager or facility owner, to explore the construction site seamlessly in time, monitor the progress of construction, assess errors and deviations, and create photorealistic architectural visualizations. These interactions are facilitated by automatic reasoning performed by our system: static and dynamic occlusions are removed automatically, rendering information is collected, and semantic selection tools help guide user input. We also demonstrate that our user-assisted SfM method outperforms existing techniques on both real-world construction data and established multi-view datasets.
\end{abstract}

\begin{CRcatlist}
\CRcat{H.5.1}{Information Interfaces and Presentation}{Multimedia Information Systems}{Artificial, Augmented, and Virtual Realities}
\CRcat{I.2.10}{Artificial Intelligence}{Vision and Scene Understanding}{Perceptual reasoning};
\end{CRcatlist}

\keywordlist

\TOGlinkslist

\copyrightspace

%-------------------------------------------------------------------------------
\section{Introduction}
\begin{figure}[t]
\includegraphics[width=.495\linewidth]{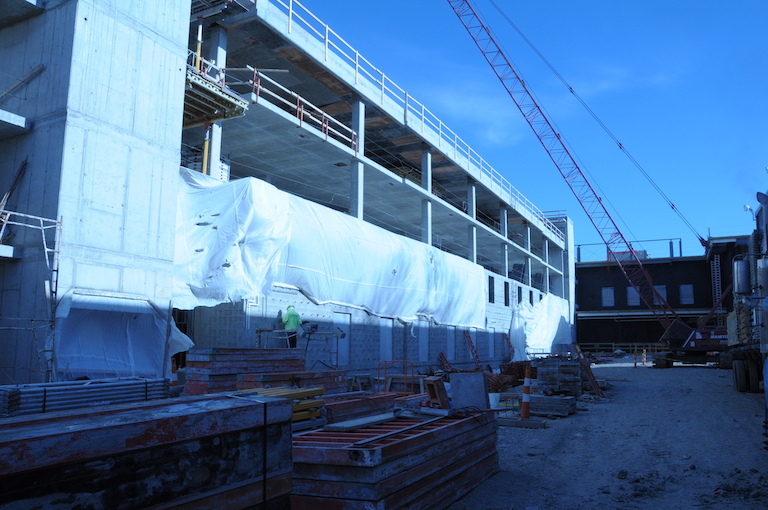}
\includegraphics[width=.495\linewidth]{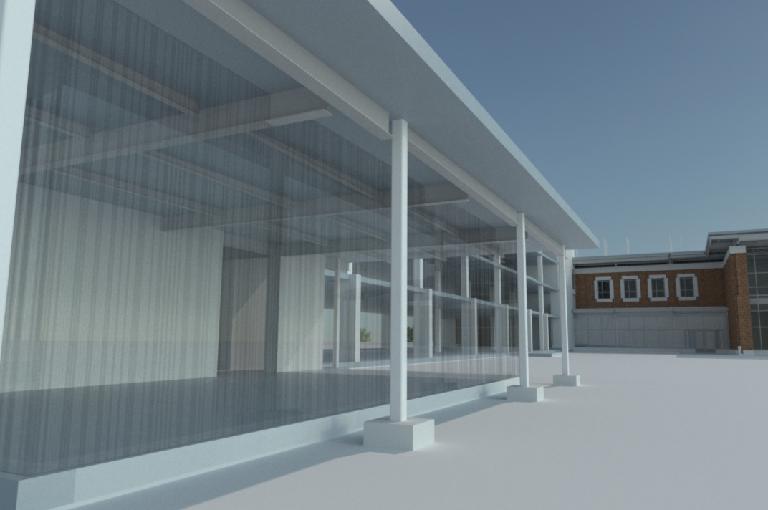}\\
\includegraphics[width=.495\linewidth]{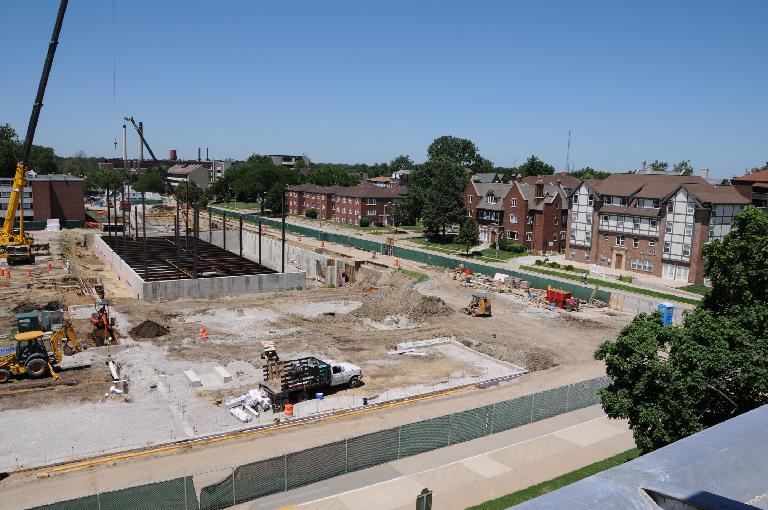}
\includegraphics[width=.495\linewidth]{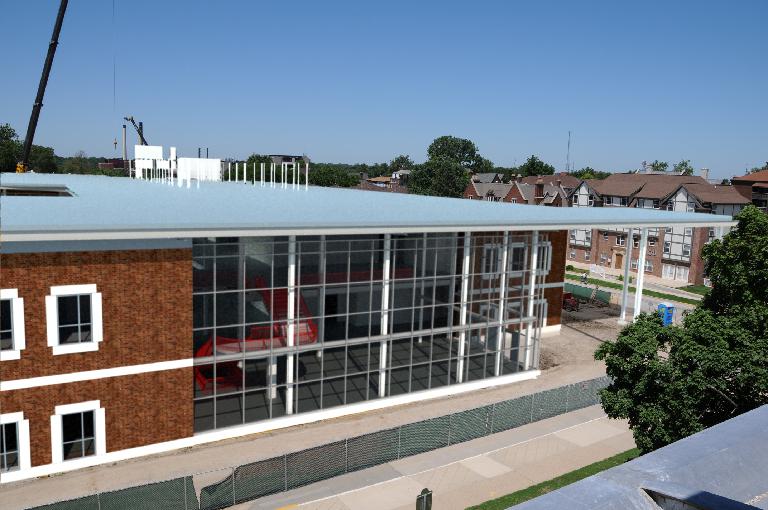}\\
\includegraphics[width=.495\linewidth]{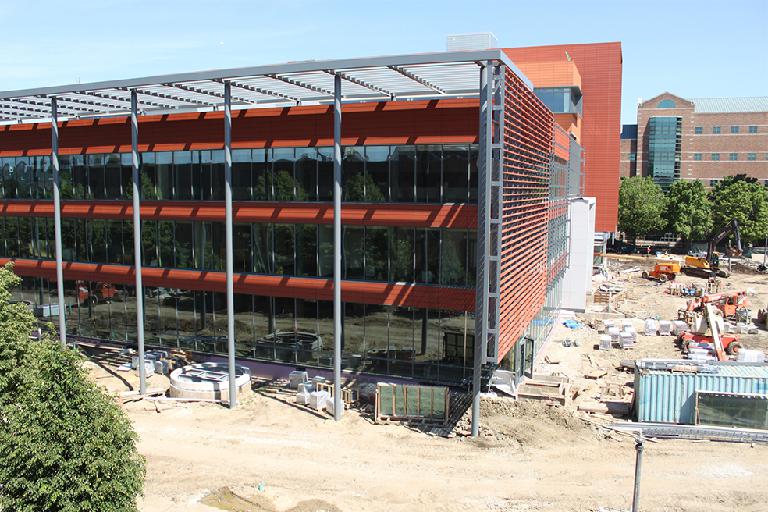}
\includegraphics[width=.495\linewidth]{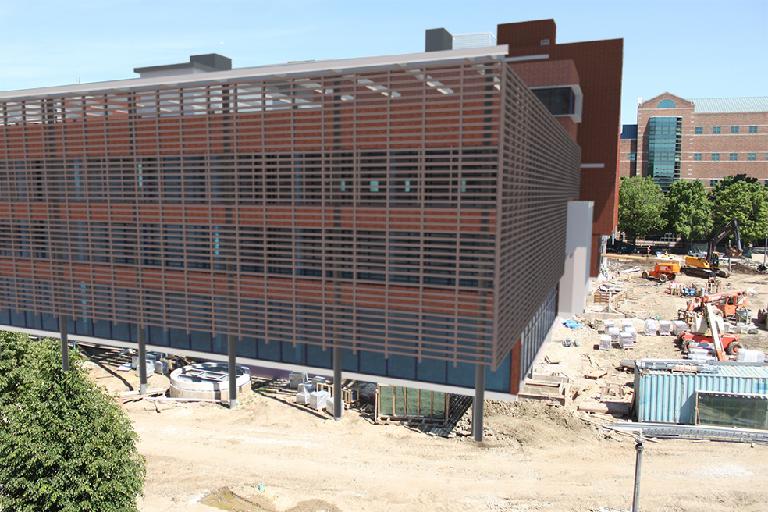}
\caption{Our software automatically renders and composites photorealistic visualizations of the building model into in-progress construction photos. Original photos on left, visualizations on right.}
\label{fig:rendering}
\end{figure}

\begin{figure}[t]
\includegraphics[width=0.495\linewidth]{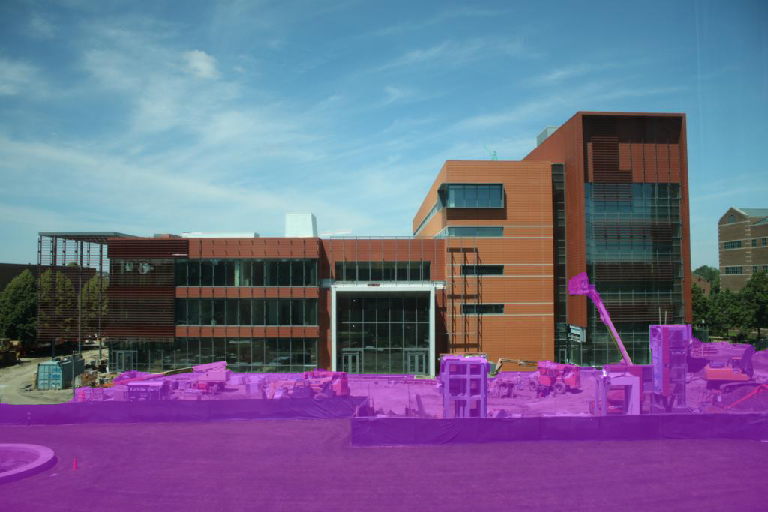}
\includegraphics[width=0.495\linewidth]{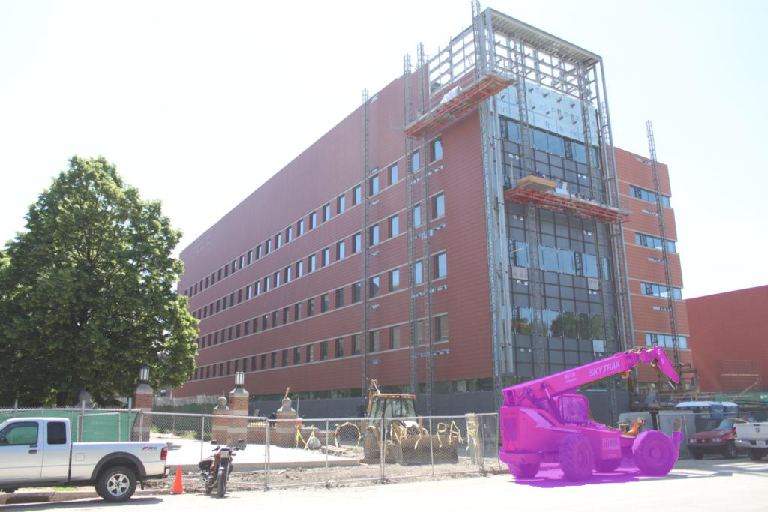}\\
\includegraphics[width=0.495\linewidth]{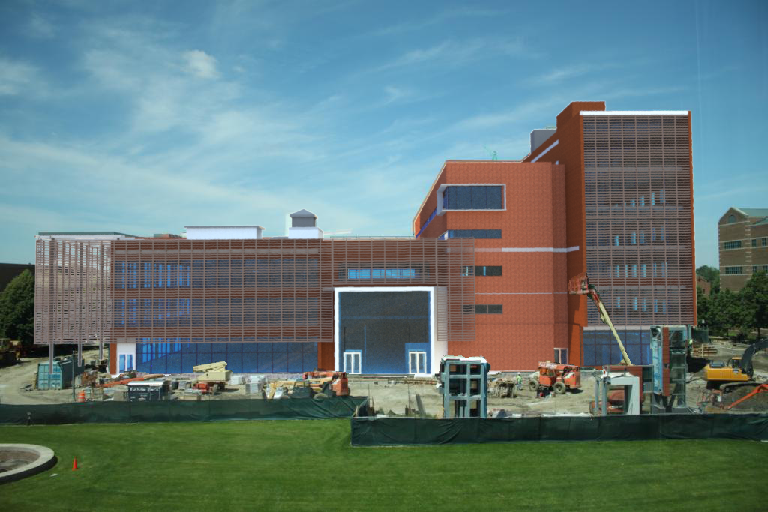}
\includegraphics[width=0.495\linewidth]{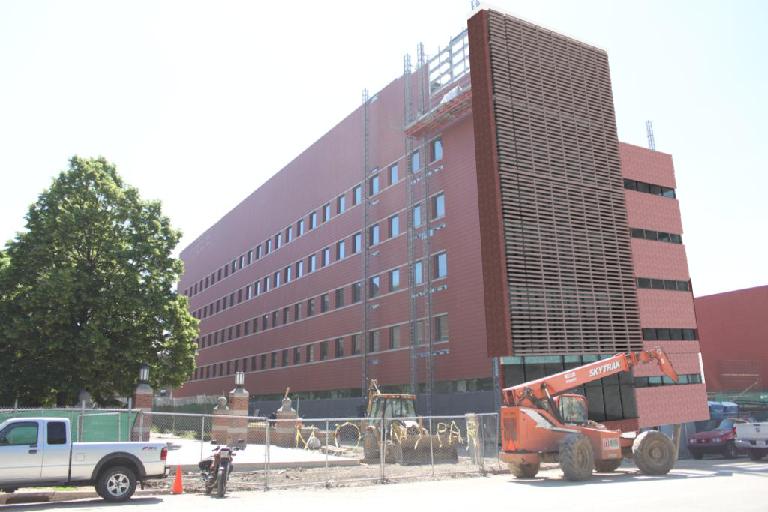}
\caption{Occlusions can be identified by analyzing the structure-from-motion point cloud and/or past imagery, allowing for occlusion removal or advanced compositing.
}
\label{fig:occlusions}
\end{figure}

\begin{figure}[t]
\includegraphics[width=\linewidth]{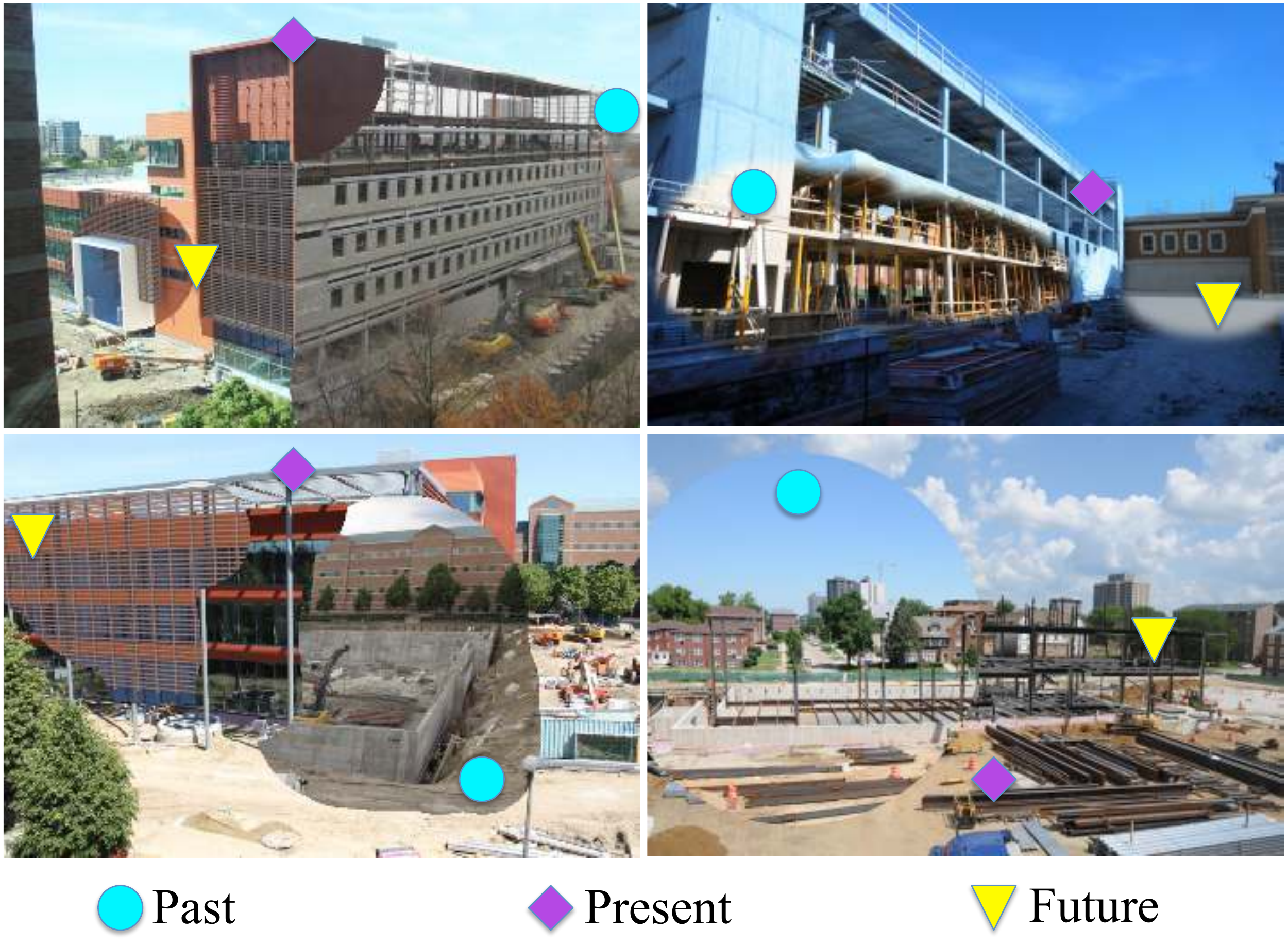}
\caption{Our system allows users to navigate image and construction models in 4D. Here, a user has selected to visualize both past and present information on each photograph.}
\label{fig:4d}
\end{figure}

%-------------------------------------------------------------------------------
On construction sites, visualization tools for comparing 3D architectural/ construction models with actual performance are an important but often unfeasible commodity for project managers~\cite{turkan2012automated}. In this paper, we develop a new, interactive method for 4D (3D+time) visualization of these models using photographs from standard mobile devices. \newtext{Our system works with unordered photo collections of any size (one picture to hundreds or more). Aligning photographs to the models to enable a suite of architectural and construction task related interactions: \vspace{-1mm}

\begin{itemize}
\item {\bf Photorealistic visualization}. Automatically create architectural renderings overlaid realistically onto photographs (Fig~\ref{fig:rendering}) and identify and segment occluding elements on the job site (e.g. construction equipment and vehicles, Fig~\ref{fig:occlusions}). \vspace{-1mm}
\item {\bf Performance monitoring}. Track the current state of construction to determine components which have been constructed late, on time, or constructed according to the building plan (or not) (Fig~\ref{fig:progress}). Annotations made on one site photo are automatically transferred to other site photos (both new and existing) for fast annotation and collaborative editing/analysis. \vspace{-1mm}
\item {\bf 4D navigation}. Selectively view portions of a photographed scene at different times (past, present and future, Fig~\ref{fig:4d}).
\end{itemize}}

\newtext{
This system is our primary contribution. We also demonstrate a new, user-assisted Structure-from-Motion method, which leverages 2D-3D point correspondences between a mesh model and one image in the collection.} We propose new objective functions for the classical point-$n$-perspective and bundle adjustment problems, and demonstrate that our SfM method outperforms existing approaches.

\subsection{Design Considerations}
\boldhead{Architectural Visualizations} A common and costly problem for designing new buildings or renovating existing facilities is misinterpretation of the building design intents. Our system allows for interactive visualizations of architectural models using photographs taken from desired viewpoints, and conveys a greater spatial awareness of a finished project. It also encouraging homeowners and facility managers to \textit{interact} with the design model, creating the ability to \textit{touch}. Building fa\c cades, architectural patterns, and materials can all be experimented with and quickly altered, and customized to individual preference from any desired viewpoint. It also promotes efficiency in professional practices by shortening the duration of design development and coordination processes.

%-------------------
%-------------------

%-------------------
%-------------------

\boldhead{Construction Visualizations} On-demand access to project information during the construction phase has a significant potential for improving decision-making during on-site activities. Visualizing 4D models with photographs provides an unprecedented opportunity for site personnel to visually interact with project documents, geo-localize potential errors or issues, and quickly disseminate this information to other users across the project. It can also facilitate field reporting and quality inspections as it allows iterations of work-in-progress and inspections to be properly logged. A time-lapse sequence of rendered images can also act as rich workflow guidelines (especially when contractors require detailed and step-by-step instructions), facilitate onsite coordination tasks, and minimize changes of requests for information from the architects. Facility owners and managers can also easily review their project at any time during the construction phase. These minimize inefficiencies that cause downtime, leading to schedule delays or cost overruns.

\boldhead{Facility Management Visualizations} The ability to illustrate what elements lay within and behind finished surfaces (e.g., a wall) and interact with them -- either through photos previously captured during the construction/renovation phase or 3D architectural model -- during the operation phase of existing facilities is of tremendous value to facility managers. Joint rendering of envisioned construction versus actual construction can facilitate inventory control tasks and simplify recordings related to repair histories.

Existing tools for addressing these needs fall into two categories: one group of tools (e.g., Studio Max, MicroStation) allow users to interactively insert 3D models into single or time-lapse photos. The second group are mobile augmented reality systems that rely on radio-frequency based location tracking, fiducial markers, or on-board sensors to track location and orientation of user and superimpose 3D models into live video streams.
%These tools are incapable of producing accurate and visually aesthetic renderings due to inaccuracies and latencies in today's commodity location-tracking technologies.
There are also challenges in storing and frequently updating large 3D models, together with relevant project information, on mobile devices.
%Also none of existing tools reason about occlusions and thus create artifacts during photo augmentation.
All these challenge frequent application of visualization tools for site monitoring purposes, and thus may minimize opportunities for detecting and communicating performance deviations before they result in schedule delays or cost overruns.

%-------------------------------------------------------------------------------
\section{Related Work}
%-------------------------------------------------------------------------------

The problem of registering large numbers of unordered ground photos, time-lapse videos, and aerial imagery with 3D architectural models has received tremendous interest in the civil engineering, computer graphics, and computer vision communities. Significant success has been reported with semi-automated systems for registering 3D architectural/construction models with time-lapsed videos~\cite{golparvar2009visualization,Kahkonen}, and using radio-frequency based location tracking or fiducial markers for augmented reality visualization of 3D CAD models for head-mounted displays~\cite{Behzadan:2005,dunston2003mixed,wang2009augmented,hammad2009distributed} and more recently commodity smartphones~\cite{cote2013live,hakkarainen2009software,irizarry2012infospot,woodward2010mixed,lee2011augmented,shin2010technology,yabuki2010collaborative}. %,zollmann2012comprehensible,Hakkarainen,wang2013augmented,schall2012smart

Among related work, D4AR modeling~\cite{JCEM2011} is the most closely related to ours. Using an unordered collection of site photos, the underlying geometrical model of a construction site is captured using a pipeline of Structure-from-Motion (SfM)~\cite{SNAVELY-IJCV08} and Multi-View Stereo~\cite{furukawa2010accurate}. By solving the similarity transformation between the 3D CAD model and the point cloud using a few user inputs, the point cloud is transformed into the CAD coordinate system, allowing the CAD models to be seen through SfM images. Using the ``traffic light'' metaphor as in~\cite{aravind2007} the photos can be augmented with color coded CAD elements.
%Our system requires less input, performs a more accurate and complete 3D reconstruction, and gives users greater control of visualization and analysis tools.
%Our system requires less input, and gives users greater control of visualization and analysis tools.

%-------------------
\begin{figure}[t]
\includegraphics[width=\linewidth]{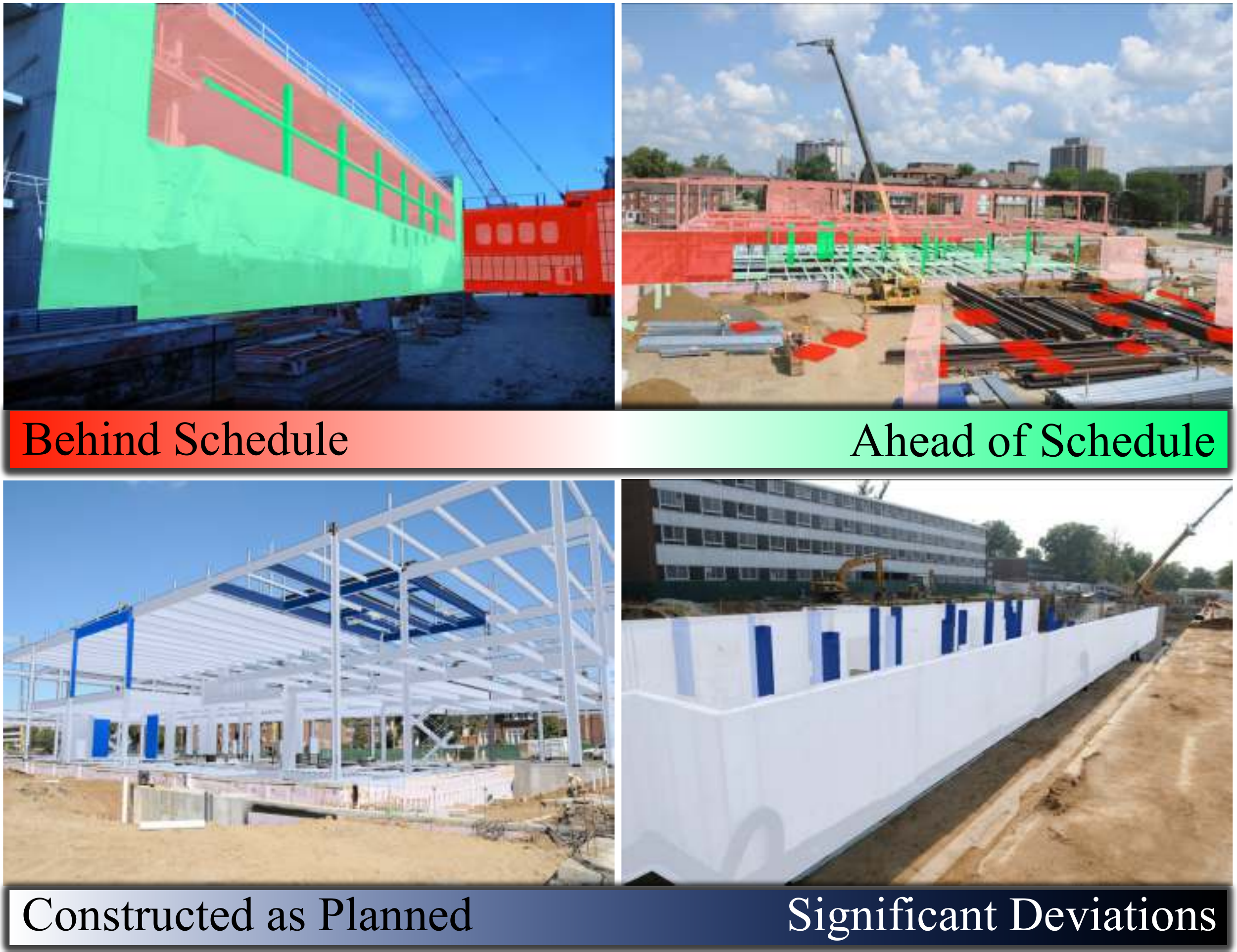}
\caption{Construction photos can be annotated using a palette of smart-selection tools, indicating which regions/building components need review or are behind or ahead of schedule. Annotations are automatically transferred among all views (including those at previous/future dates), allowing field engineers and project managers to collaborate and save time in creasing visualizations.
\vspace{-1mm}
}
\label{fig:progress}
\end{figure}
%-------------------

%-------------------
%-------------------
Other works on construction site visualization techniques include manual occlusion management~\cite{zollman_occlusion} and ``x-ray'' metaphors~\cite{zollmann_xray}. While these methods are intended for augmented reality / head-mounted displays, our work improves on these techniques by automatically estimating occlusions and allowing for more control in the visualization process.

Our semi-automated method for registering 3D models to photographs is inspired by past work on registration~\cite{franken2005minimizing,PintusFast}, architectural reconstruction~\cite{Wan201214,Werner:2002}, and automatic camera pose estimation/calibration from unordered photo collections~\cite{SNAVELY-IJCV08,SGSS-siggraph08}. Such methods, including ours, are known as ``incremental SfM'' (adding one or a few photo(s) at time), and recent methods demonstrate improvements by solving the SfM problem at once~\cite{disco12pami}. Our method is also related to the user-guided SfM method of Dellepiane et al.~\shortcite{semiauto_sfm}, although the inputs and goals our systems are different (input comes in the form of taking additional photos to guide reconstruction).

Several methods exist for aligning 3D models automatically in photographs~\cite{Huttenlocher_reg,Lowe_reg}, and more recently Russell et al.~\shortcite{Russell-3DRR11} demonstrate a technique for automatically aligning 3D models in paintings. However, such methods are not suitable for construction sites since the 3D model and photographs rarely correspond in appearance (e.g., construction photos contain many occlusions, missing or added elements not present in the 3D model; the level of detail in 3D model may not match the level of detail in actual elements on site). Although unrelated to construction, Bae et al.~\shortcite{Bae:2010} demonstrated a method for merging modern and historical photos from similar viewpoints to allow for temporal navigation.

Many techniques exist for 3D architectural and mesh modeling from photographs~\cite{videotrace,sinhaArch,ImageRemodel,xu_sig11,Debevec:facade}. Our method instead relies on an existing, semantic 3D CAD model (known as a building information model, BIM). BIM are widely available as majority of building jobs require such models prior to construction.

Several methods leverage 3D models for photo editing, rendering, and visualization~\cite{DeepPhoto,city4d,view4d}. Most similar to our technique, Schindler and Dellaert in particular describe a method for navigating historical photo collections, including 4D interactions. Distinct from other techniques, our method utilizes a semantic 3D model to enable 4D visualizations from arbitrary viewpoints (provided enough images exist), as well as tools for collaborative analysis and visualization, occlusion identification, and photorealistic rendering.

%-------------------
\begin{figure*}[ht]
\includegraphics[width=\linewidth]{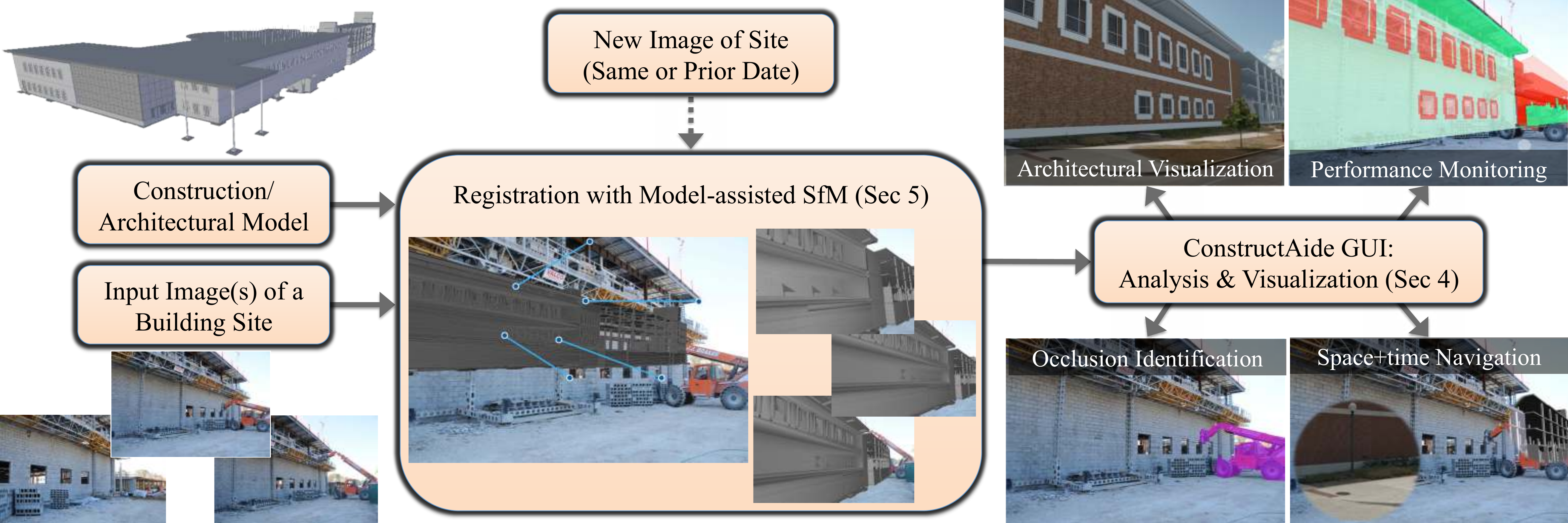}
\caption{Method overview. Our system takes as input a 3D model and one or more photos of a construction site. The model and photos are  aligned using our Model-assisted SfM approach: one image is registered by specifying 2D-3D correspondences, and other images are then registered automatically. Leveraging aligned photo-mesh data, we extract automatic estimates of occlusion, rendering information, and selection aids. Our interface then allows users to explore the model and photo data in 4D and create informative/ photorealistic visualizations.}
\label{fig:overview}
\end{figure*}
%-------------------

%-------------------------------------------------------------------------------
\section{System Overview}
%-------------------------------------------------------------------------------

Our approach, outlined in Figure~\ref{fig:overview}, takes advantage of a small amount of user input to register all photos with the underlying 3D architectural/construction model. We only ask the user to specify a few correspondences between an image and the underlying 3D model, providing a registration between the model and the photo. Our system then registers other images automatically using our proposed Structure-from-Motion (SfM) formulation (Sec~\ref{sec:MeshSFM}). New photos of the same site -- taken at either an earlier or later date -- can also be registered with no additional interaction.

%We also use the same information to register photos previously captured into the same coordinate system. Leveraging priors in the underlying architectural model, our method is suitable for realistically rendering the architectural/construction models with different levels of detail. We also introduce new functionalities to slice into the geometry and time to create time-lapsed views.

Once the 3D model is registered with the photograph(s), we preprocess the images to estimate timelapses from unordered photo sets (Sec~\ref{sec:4D:timelapse}), static and dynamic occlusions (Sec~\ref{sec:4D:occ}), and light/material models for rendering (Sec~\ref{sec:4D:buildinginfo}). Our user interface (Sec~\ref{sec:4D:GUI}) provides simple visualization metaphors that enable a user to interact with and explore the rich temporal data from the photo sets and architectural/construction models. For example, a user can quickly select elements from the photographs at any point in time, hide/show the elements, visualize the construction progress or analyze errors. Annotations and visualizations are automatically transferred across views, allowing for real-time, collaborative analysis and viewing. Finally, photorealistic architectural renderings can be produced without the user ever using CAD, 3D modeling or rendering software: the model is rendered using the extracted material/lighting model and composited back into the photograph automatically.

%----------------------------------------------
\newtext{
\subsection{Assumptions and Input Requirements}
ConstructAide relies on accurate BIM (semantically rich CAD models), including complete, up-to-date mesh models and scheduling information.} \newnewtext{We require BIM as these allow us to easily hide/show parts of the 3D model based on construction schedules, allowing for easy component selection and annotation at any phase of the build process. Furthermore, the information contained in BIM allows us to automatically produce photorealistic renderings composited onto in-progress construction photographs.}
\newtext{In order to best traverse and visualize the construction process using our system's tools, a spatially and temporally dense set of photographs is required. However, results can still be achieved with fewer images; registration, photorealistic rendering, and performance monitoring can be done with a single photograph, and occlusion identification and 4D navigation are possible with two or more photos (given the view location/orientations are similar). As more photographs are collected and added, our system enables more automation: Mesh-SfM automatically registers new images, annotations can be transferred across views, and occlusion estimates improve.
}
\newnewtext{The photo collections in this paper typically contain 10-25 photos, which we found is sufficient to obtain reasonable results.}

%-------------------------------------------------------------------------------
\section{ConstructAide System}
%-------------------------------------------------------------------------------
\label{sec:4D}
Our interface requires as input one or more photos of the job site, a 3D building model, and an accurate registration of the model to each of the photos. In this section, we assume that the registration process is complete (Sec~\ref{sec:MeshSFM} describes our registration approach, but other approaches or ground truth data could be used if available). Our system allows users to virtually explore the job site in both space and time, analyze and assess job progress, and create informative visualizations for the construction team. We first preprocess the input (images and 3D model) as in Sec~\ref{sec:4D:preprocess}, and then the user can begin virtually interacting with the construction site (Sec~\ref{sec:4D:GUI}).

%--------------------------------------
\subsection{Preprocessing}
\label{sec:4D:preprocess}
To enable more efficient interactions, we first process the registered data to extract information useful for selection, visualization, and rendering. For example, converting unordered collections into time-lapse data, identifying and removing occlusions, and extracting rendering information from building models enable users to navigate and visualize data with ease, allowing for valuable job-site visualizations to be created in minutes.

%--------------------------------------
\subsubsection{Converting Unordered Image Sets into Time-lapses}
\label{sec:4D:timelapse}
The first step in this process is, for each image, to identify other images that were taken from roughly the same viewpoint, determined by how well a single homography can model matched features in every pair of images. We have already computed this data for registering the construction models to the photos (described in Sec~\ref{sec:MeshSFM}), and there is no need to recompute homographic transformations. Once similar-viewpoint pairs are identified, the homography is used to transform one image into the other's view; we do this at each camera location and for all nearby viewpoints, resulting in pixel-aligned temporal information. If no nearby viewpoints are found, this image cannot be traversed temporally in 2D (however, the registered 4D mesh can still be traversed).

%-------------------
\begin{figure}[t]
\includegraphics[width=0.49\linewidth,clip=true,trim=0 0 0 300]{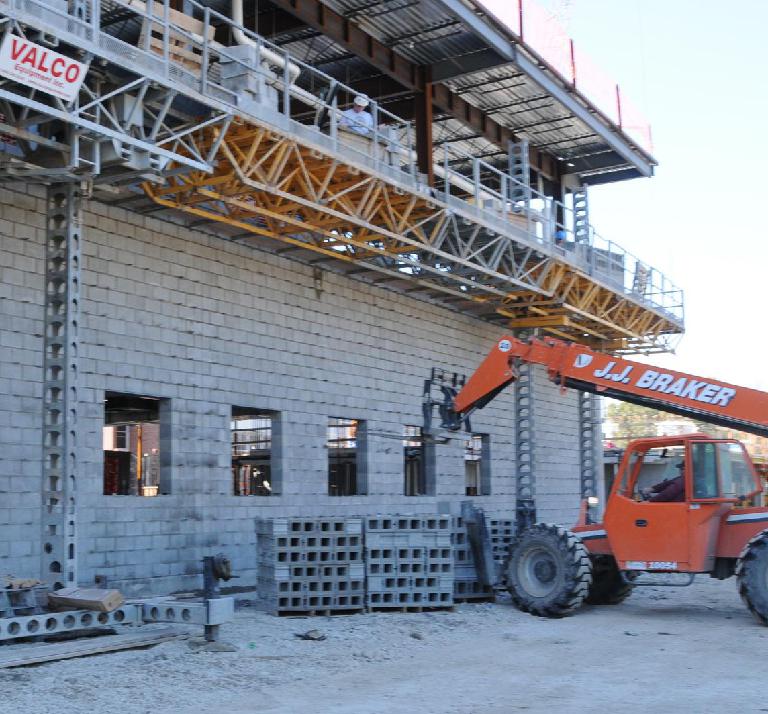}
\begin{overpic}[width=0.49\linewidth,clip=true,trim=0 0 0 300]{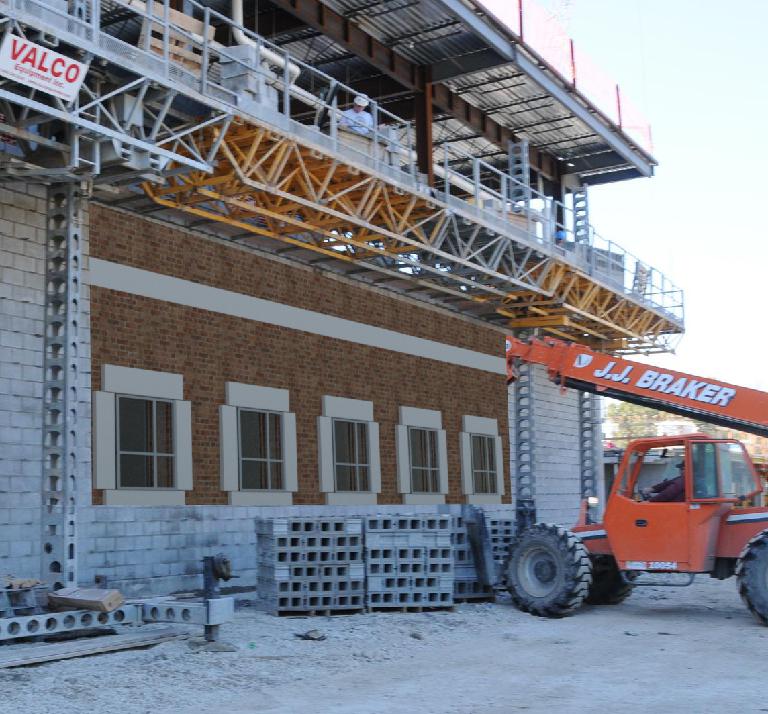}
\setlength\fboxsep{0pt}
\setlength\fboxrule{0pt}
\put(92,2){\fbox{\includegraphics[width=0.1\linewidth]{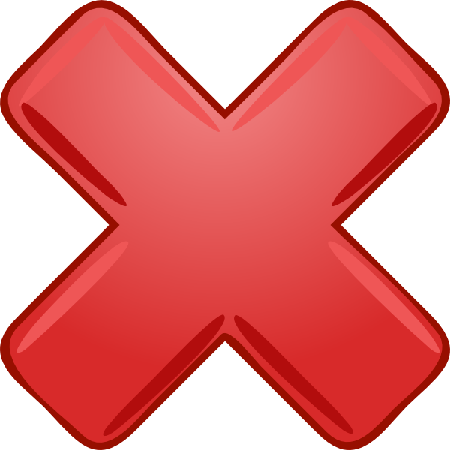}}}
\end{overpic}
\\
\includegraphics[width=0.49\linewidth,clip=true,trim=0 0 0 300]{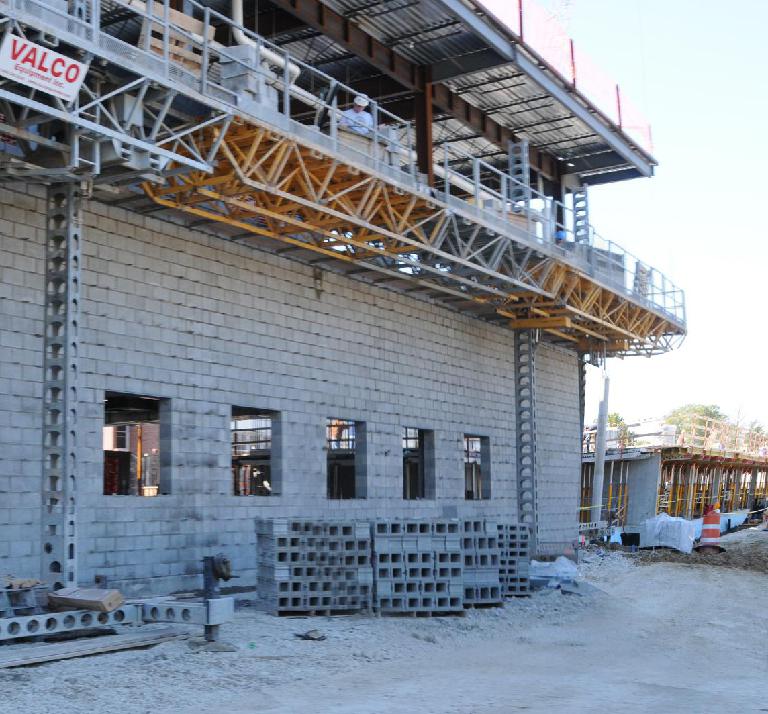}
\begin{overpic}[width=0.49\linewidth,clip=true,trim=0 0 0 300]{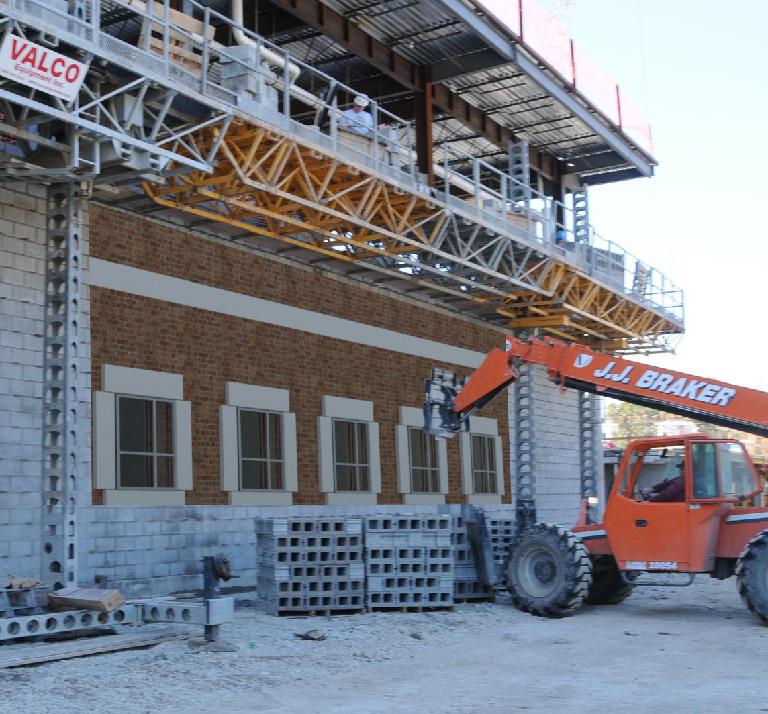}
\setlength\fboxsep{0pt}
\setlength\fboxrule{0pt}
\put(90,0){\fbox{\includegraphics[width=0.12\linewidth]{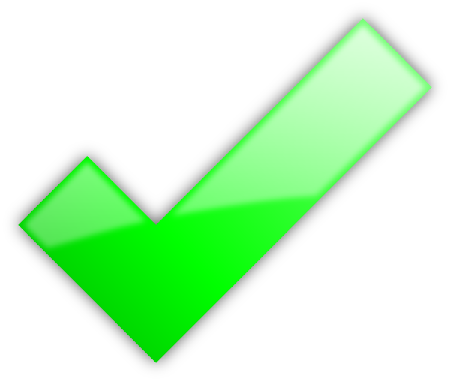}}}
\end{overpic}
\caption{Our method reasons about dynamic occlusions (such as the red telehandler pictured in the top left) by sampling similar viewpoints at different times so that depth layers are not confused in visualizations (top right). A clean background image is computed automatically (bottom left), and an occlusion mask is created by comparing the original image with the background image, allowing for building elements to be visualized with proper occlusion (bottom right).}
\label{fig:dynamicOcc}
\end{figure}
%-------------------

%-------------------
\begin{figure*}[!ht]
\setlength\fboxsep{0pt}
\setlength\fboxrule{0pt}
\includegraphics[width=0.16\linewidth,clip=true,trim=0 0 0 30]{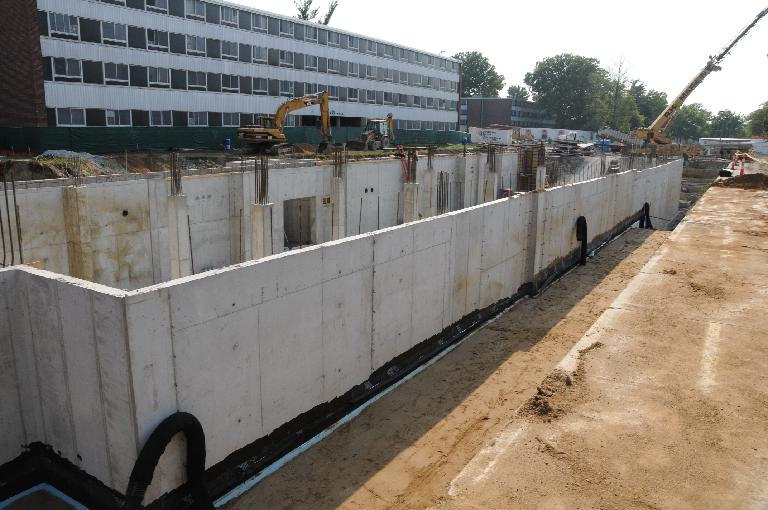}
\begin{overpic}[width=0.16\linewidth,clip=true,trim=0 0 0 30]{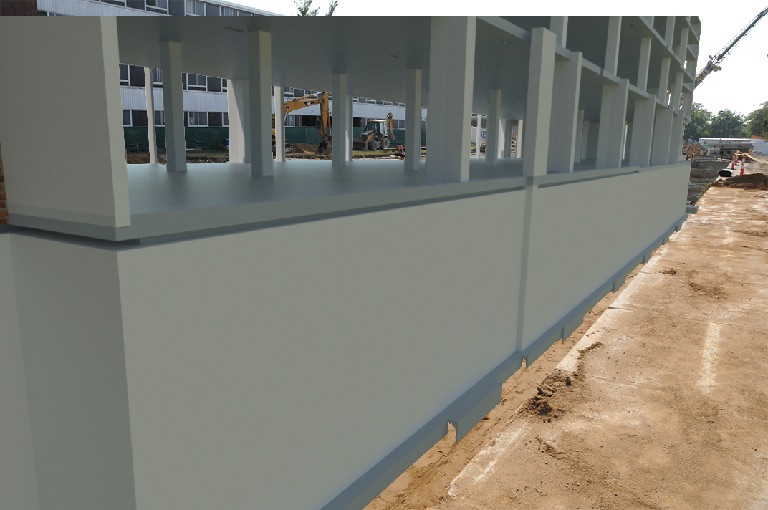}
\put(62,2){\fbox{\includegraphics[width=0.03\linewidth]{fig/dynamicOcc/x.png}}}
\end{overpic}
\begin{overpic}[width=0.16\linewidth,clip=true,trim=0 0 0 30]{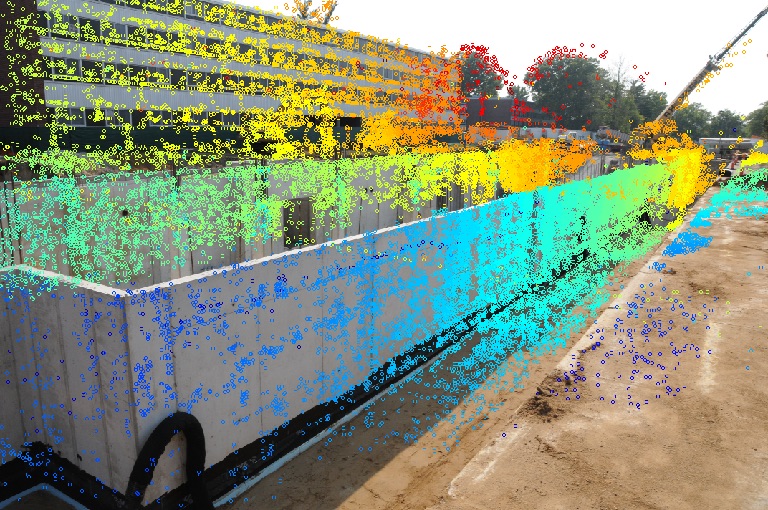}
\put(20,0){\fbox{\includegraphics[width=0.12\linewidth]{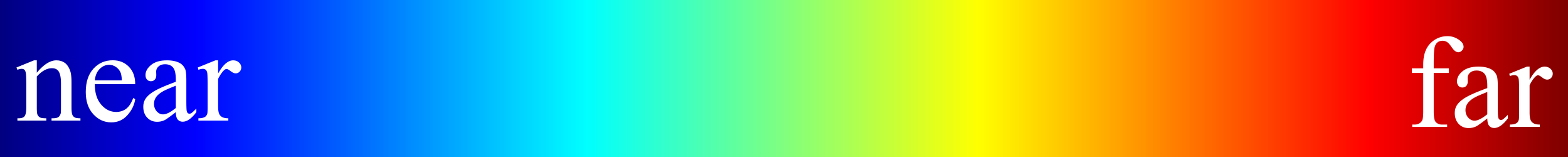}}}
\end{overpic}
\includegraphics[width=0.16\linewidth,clip=true,trim=0 0 0 30]{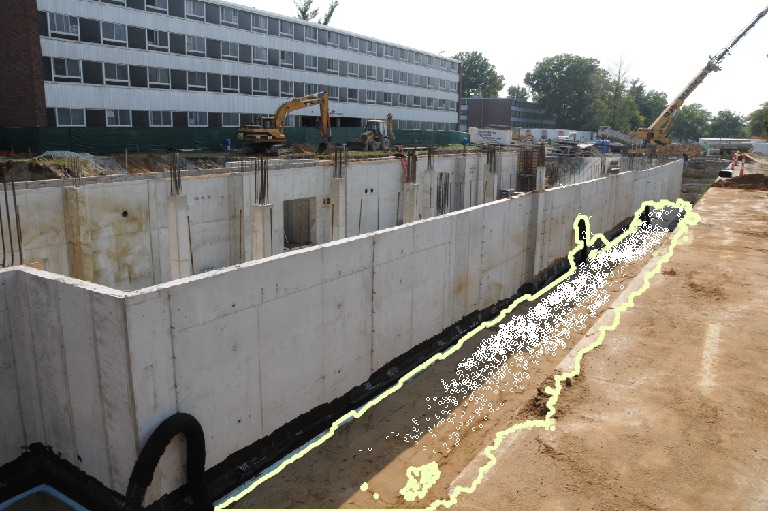}
\hfill
\begin{overpic}[width=0.16\linewidth,clip=true,trim=0 0 0 30]{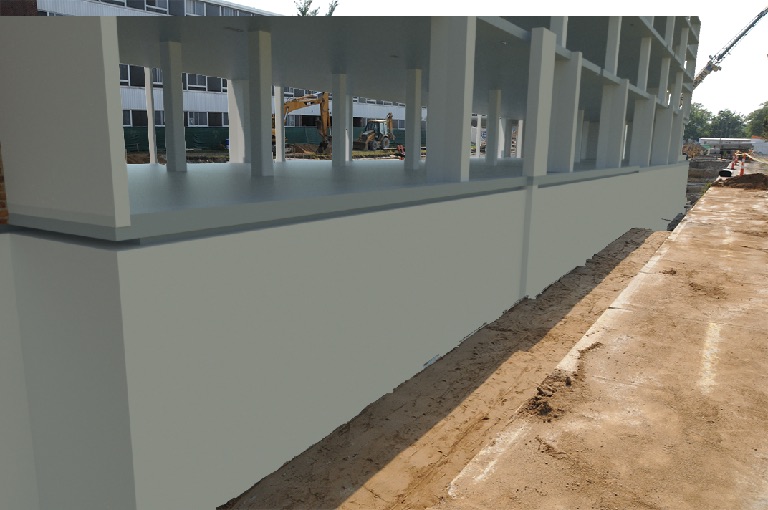}
\put(60,1){\fbox{\includegraphics[width=0.04\linewidth]{fig/dynamicOcc/check.png}}}
\end{overpic}
\begin{overpic}[width=0.16\linewidth,clip=true,trim=0 0 0 30]{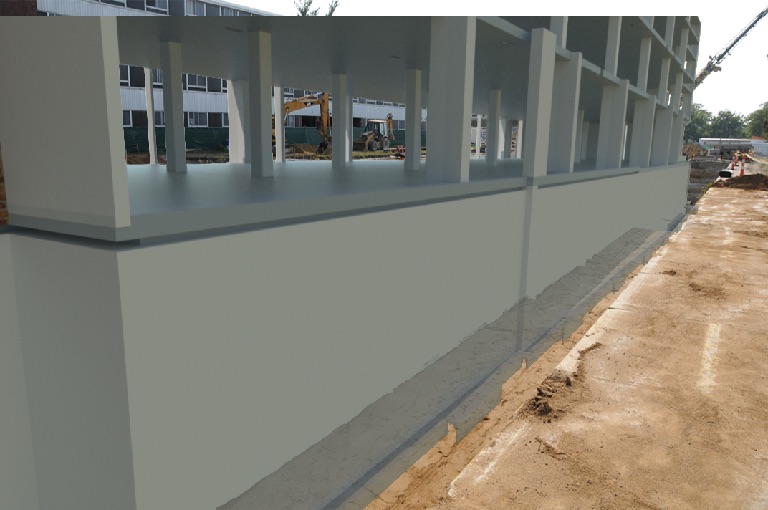}
\put(60,1){\fbox{\includegraphics[width=0.04\linewidth]{fig/dynamicOcc/check.png}}}
\end{overpic}
\\
\begin{minipage}{0.16\linewidth}\centerline{(a)}\end{minipage}
\begin{minipage}{0.16\linewidth}\centerline{(b)}\end{minipage}
\begin{minipage}{0.16\linewidth}\centerline{(c)}\end{minipage}
\begin{minipage}{0.16\linewidth}\centerline{(d)}\end{minipage}
\hfill
\begin{minipage}{0.16\linewidth}\centerline{(e)}\end{minipage}
\begin{minipage}{0.16\linewidth}\centerline{(f)}\end{minipage}
\caption{Our system attempts to detect and static occlusions (i.e. the basement of this building model is hidden by the ground in (a) by comparing the mesh model (b) the point cloud estimated with SfM (c). 3D points that are measured to be in front of the model (see text for details) are then propagated and smoothed based on image appearance, resulting in an occlusion mask (d). The occlusion mask can be used to properly hide basement elements (e), or create an x-ray type visualization (f).}
\label{fig:staticOcc}
\end{figure*}
%-------------------

%-------------------
\begin{figure*}[t!]
\includegraphics[width=.16\linewidth]{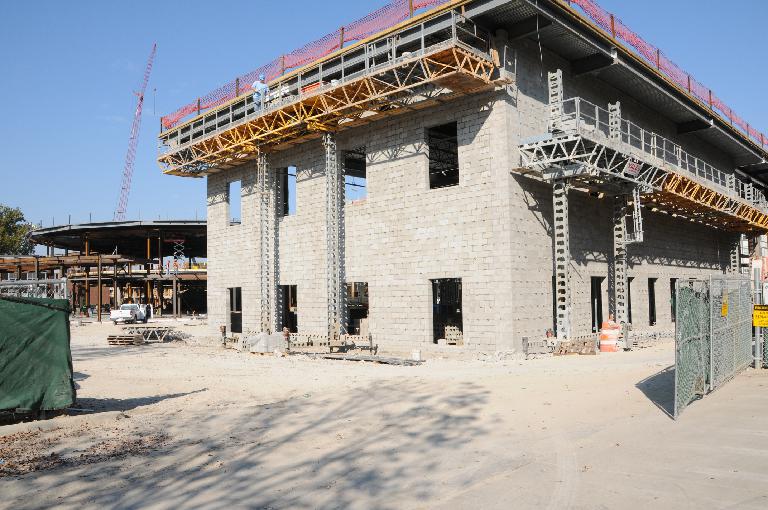}
\includegraphics[width=.16\linewidth]{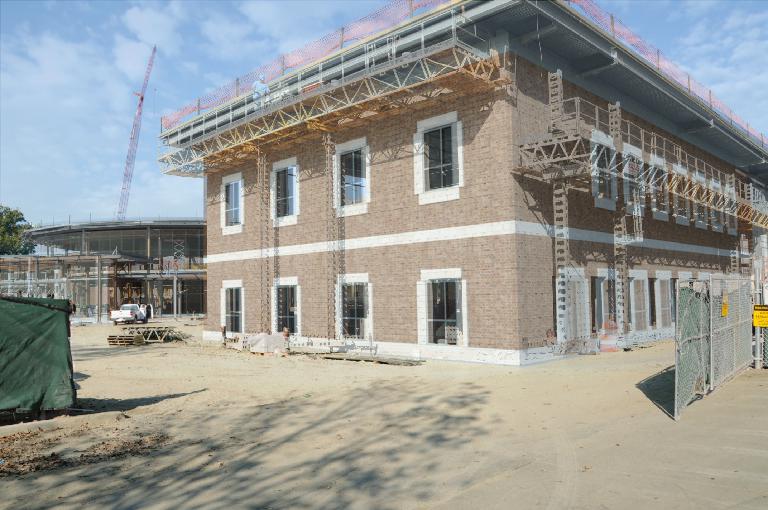}
\includegraphics[width=.16\linewidth]{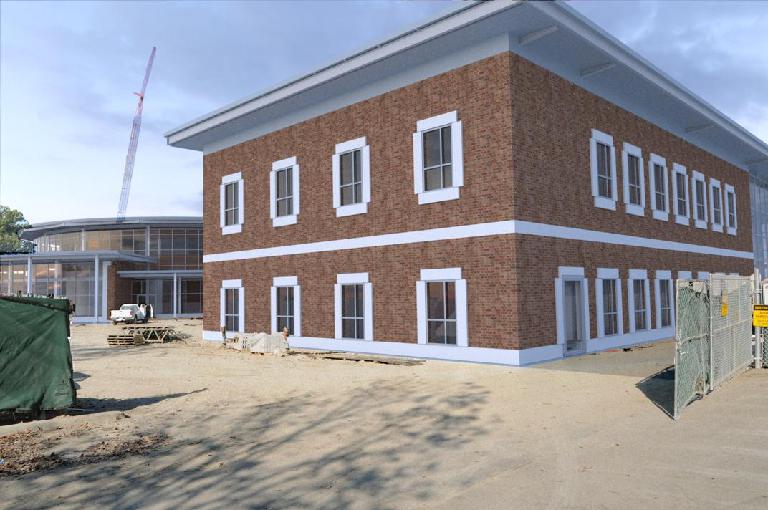}
\includegraphics[width=.16\linewidth]{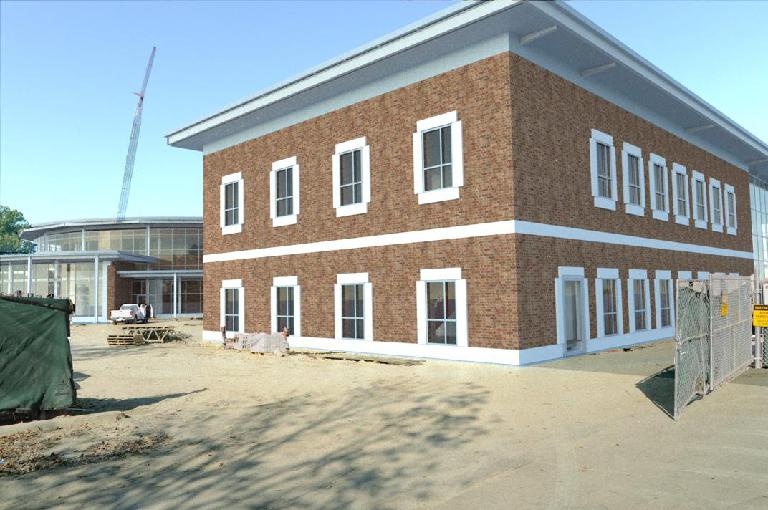}
\includegraphics[width=.16\linewidth]{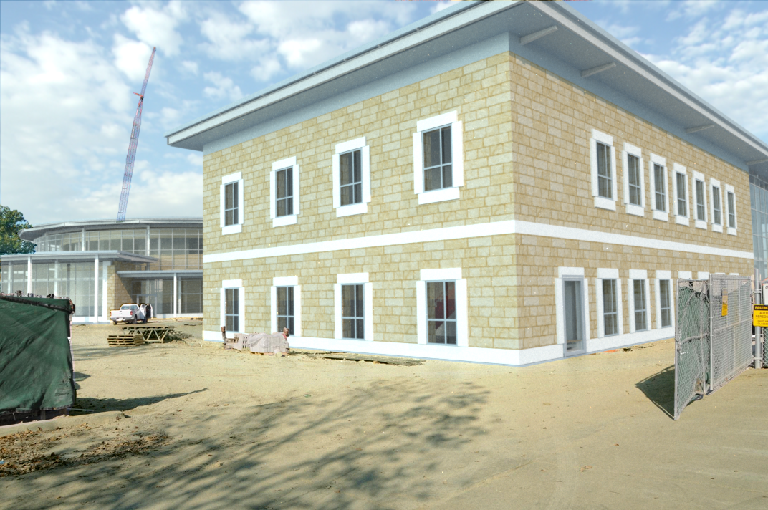}
\includegraphics[width=.16\linewidth]{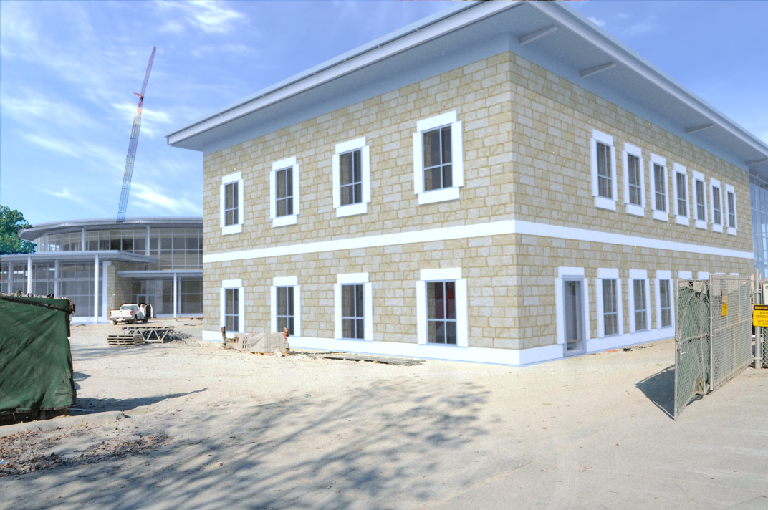}
\caption{Our system can be used to create photorealistic architectural visualizations automatically rendered {\it into} the photograph without the use or knowledge of any CAD, modeling, or rendering software. Here, we show a construction image, followed by a blended architectural render, and four different material / lighting choices for the scene. Occlusion information is computed automatically; errors and other objects can be corrected/added using our efficient selection tools (e.g., bottom row; the truck and crane were added manually, and we make no attempt to modify the shadow cast by the fence). Time lapses and changes to materials can be rendered with ease by swapping out preset HDRI light sources and materials.}
\label{fig:timelapse}
\end{figure*}
%-------------------

%--------------------------------------
\subsubsection{Occlusion Identification}
\label{sec:4D:occ}
We attempt to automatically identify troublesome occlusions that can lead to unappealing visualizations. For example, a truck may be idle temporarily in front of a fa\c cade (dynamic occlusion), or certain building components may be built beneath the ground or behind other non-building structures (static occlusions) -- see Figs~\ref{fig:dynamicOcc} and ~\ref{fig:staticOcc}. Such occlusions can be a nuisance when creating visualizations, and manually removing them may take time and expertise.

We handle the two types of occlusion (dynamic: moving equipment and workers; static: immobile elements blocking camera's field of view) separately. For dynamic occlusions, we assume that the occluding object is only in place temporarily, and thus that it does not occupy the same pixels in a majority of the aligned time lapse data (computed in Sec~\ref{sec:4D:timelapse}). We then find the ``background'' image by computing the per-pixel median of the time lapse (disregarding the image containing the occlusion); if our assumption holds, the moving object will be removed. To identify the pixels of the dynamic object, we compute the squared pixel-wise difference (in HSV) between the original image and the background, smooth the result with the cross-bilateral filter~\cite{cross_bf}, and threshold the smoothed result, keeping pixels greater than 0.05 in any channel. Fig~\ref{fig:dynamicOcc} demonstrates this process. As few as one additional image can suffice for detecting dynamic occlusions, but more can be used if available.

For static occlusions, we attempt to identify pixels in an image which are spatially in front of the 3D model, e.g., a fence might block a fa\c cade, or the ground may occlude the model's basement. Our idea is to make use of the 3D model and the sparse set of 3D points computed during our SfM procedure (Sec~\ref{sec:MeshSFM}). For each of these 3D points $p$ project onto the 3D model, we predict whether or not this point is in front of the model by evaluating the following heuristic:

\begin{equation}
[p - p_{\text{model}} > 0.3] \vee [\cos^{-1}(n(p)^T n(p_{\text{model}})) > \pi/6],
\end{equation}

where $p_{model}$ is the 3D location corresponding to the point on the mesh $p$ projects to, and $n(p)$ calculates the surface normal at $p$ (estimated using nearby points in the point cloud). In other words, if $p$ is closer to the camera by more than 0.3m, or normals differ by more than $30^\circ$, we assume the mesh must be occluded at this pixel. The normal criterion ensures that occluding points within the 0.3m threshold are oriented properly (otherwise they belong to the ``occlusion set'').

Since our point cloud is sparse, the binary occlusion predictions will be too. To obtain a dense occlusion mask, we flood superpixels (computed using SLIC~\cite{Achanta:ijcv:12,vedaldi08vlfeat}) with the sparse occlusion estimates (if a superpixel contains an occluded pixel, it becomes part of the occlusion mask); finally we smooth this mask using a cross-bilateral filter. Our approach is shown in Fig~\ref{fig:staticOcc}.

In the event of failure either due to not enough images / triangulated points or misestimation, the user can correct errors using selection and editing tools in our interface.

%--------------------------------------
\subsubsection{Utilizing Other Building Information}
\label{sec:4D:buildinginfo}
Architectural and construction models -- commonly known as Building Information Models (BIM) -- contain rich semantic information about element interconnectivity and materials. We leverage these in our interface to improve the user's experience. Building elements are clustered by \textit{primitive}, \textit{group}, and \textit{type} to accelerate selection in the photograph, scheduling info is used to create ``snapshots'' of the model's geometry at various points in time, building element material names are used to generate renderable, computer graphics materials, and GPS coordinates are used to acquire sun position (e.g. using publicly available lookup tables \url{http://aa.usno.navy.mil/data/docs/AltAz.php}).

%--------------------------------------
\subsection{User Interface}
\label{sec:4D:GUI}
Now that the meshes (from building information model) and photos are aligned and visualization tools have been prepared, a user can interact with our system using a simple user interface. Selections in an image can be made by one of many unique ``marquee'' tools: 3D building elements can be selected individually as well as grouped by type or material, and individual faces/primitives can also be selected. These semantic tools accompany standard selection tools (lasso, brush, etc; see supplemental video). Occlusion masks obtained during our occlusion identification step are also used to create grouped pixel regions for efficiently selecting such regions. Once a selection is made in the image, the user can perform the following functions:

\begin{itemize}

\item \textbf{Photorealistic Visualization:} A user can specify a subset of the underlying mesh model (using our selection tools), and seamlessly render visible/selected mesh components into the image. Geometry, lighting, and materials are known in advance (as in Sec~\ref{sec:4D:buildinginfo}), so the model can be rendered with no user interaction, and composited back into the photo using the technique of Karsch et al.~\shortcite{Karsch:SA2011}. We demonstrate a rendered result In Fig~\ref{fig:timelapse}, we demonstrate a time-lapsed architectural visualization created with our software. Architectural rendering is performed using LuxRender (\url{http://www.luxrender.net/}), and preview rendering is done in our OpenGL interface.

\item \textbf{Performance Monitoring:} Based on scheduling data and the progress of construction visible in the image(s), a user can assess the progress of a region in the image by adding annotations. A color label can be given to indicate whether the component was built ahead of schedule (green), on time (semi-transparent white), or behind schedule (red), as shown in Fig~\ref{fig:progress}. Annotating deviations in the builing process is also possible; darker blue overlays indicate components have not been built according to plan or need review. Any visualizations and notes added to an image are propagated to all other registered views/images, allowing for efficient annotation and real-time, collaborative analysis.

\item \textbf{4D Navigation:} Users can scroll through both the spatial and temporal extent of the photo collection. For navigating in time, a user can selectively peer forward or backward in time, revealing past image data or future renderings of the 3D model (in the region selected by the user).
\end{itemize}

For a demonstration of our GUI tools, see our supplemental video.

%\boldhead{Technical details} Our SfM and bundle adjustment techniques are implemented in MATLAB and not optimized for efficiency; as such registration can take seconds (a few images) to tens of minutes (large image collections). It is well known how to optimize such algorithms~\cite{vsfm2}, and we leave this as future work.

%\boldhead{Data Requirements}
Construction site imagery typically comes from a few static/mounted cameras that record time-lapses or from project participants photographing the site at regular intervals (e.g. tens to hundreds of photos every few weeks). Thus, data is typically dense temporally, but the photos usually are not spatially dense and as such have wide baselines.  Our system and interface can work with any sampling of photographs, but temporal navigation and occlusion detection is more compelling with a dense time-sampling of images.

%--------------------------------------
\subsection{Domain Expert Evaluation}
We interviewed five domain experts\footnote{All participants have advanced degrees in construction monitoring and have worked for 2-8 years as field engineers for numerous construction sites.} with experience in construction progress monitoring and assessment. All subjects were males between the ages of 24 and 38, and no compensation was given.

In preparation to the interview, we asked participants about existing solutions for progress monitoring. All subjects responded that it is current practice to simply take notes and photographs (e.g. on mobile devices) and hand-annotate these with progress and financial information. Three participants used the words ``subjective'' or ``unreliable'' to describe this process; one participant noted that these methods are ``highly subjective since the results are based on [one field engineer's] judgement and experience.'' All but one participant mentioned that current practices are tedious and are not scalable.

Following the pre-interview, we demonstrated ConstructAide to each subject. Four of the five subjects mentioned that smart-selection tools and sharing annotations would reduce the bottlenecks of current techniques, allowing for faster and more accurate on-site inspections. Subjects generally felt that the 4D navigation feature was most useful, e.g. ``space-time navigation is very useful for change management and  assessing/verifying contractor claims. You need as-built evidence for claims.'' However, three of the participants also noted that assessing building deviations in a photographical interface may not be as useful for certain tasks since ``millimeter accuracy may be required, and not all approved building changes are reflected in the BIM.''
%One subject also had concerns about incorporating security features into ConstructAide.

Participants described further automation as desired features, e.g. ``automatic progress and error assessment using the BIM and pictures.'' One user described his ideal progress monitoring tool during the pre-interview: ``In a perfect world, field reporting would automatically synch in the cloud and be accessible anywhere using smartphones. Tools would be easy to use, objective, and based on quantitative measures.''

%-------------------
\begin{figure}
\centerline{User study questionnaire average responses}
\includegraphics[width=\linewidth]{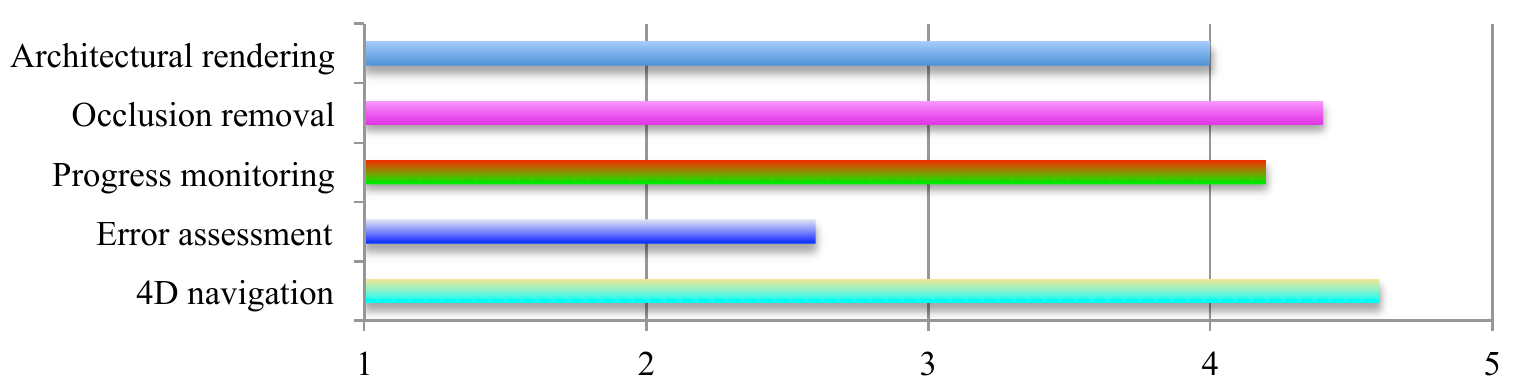}
\caption{Responses from the exit questionnaire of our user study averaged over all subjects. Each question posited the usefulness of a given feature using a Likert scale, e.g. ``Rate the usefulness of the 4D navigation tool.  (1 = poor, 5 = excellent)''}
\label{fig:studyresponses}
\end{figure}
%-------------------

%-------------------
\begin{figure}[t]
\includegraphics[width=.325\linewidth]{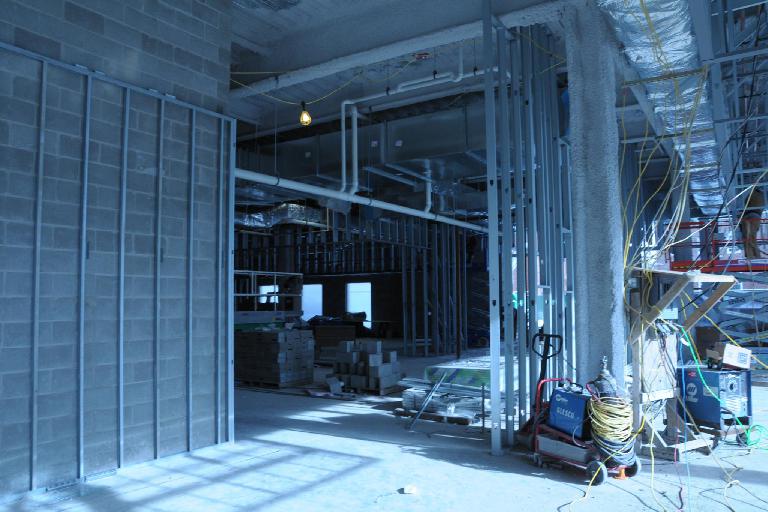}
\includegraphics[width=.325\linewidth]{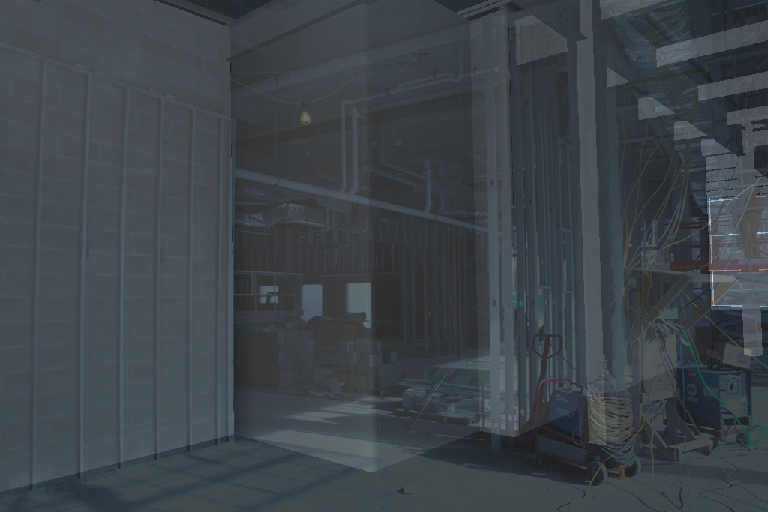}
\includegraphics[width=.325\linewidth]{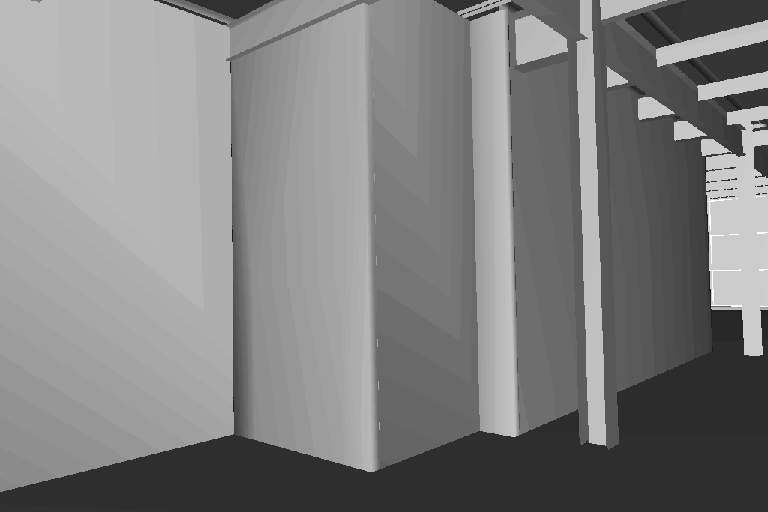}\\
\includegraphics[width=.325\linewidth,clip=true, trim=0 100 0 0]{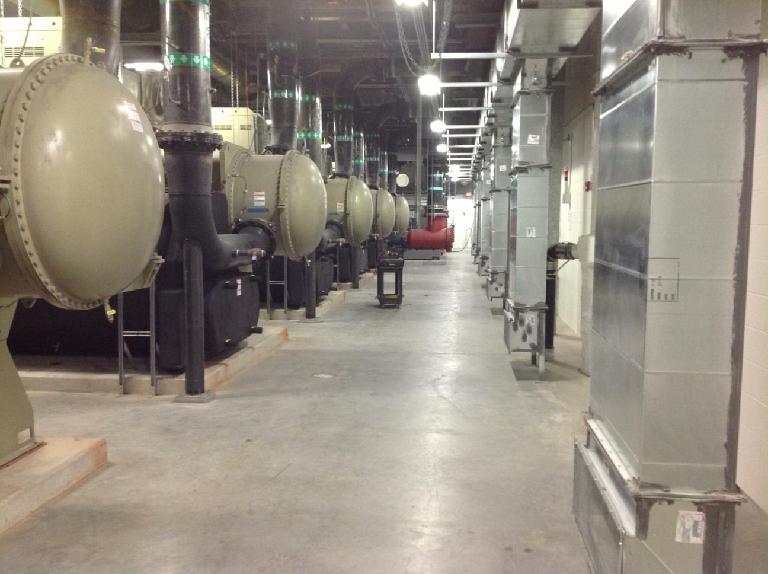}
\includegraphics[width=.325\linewidth,clip=true, trim=0 100 0 0]{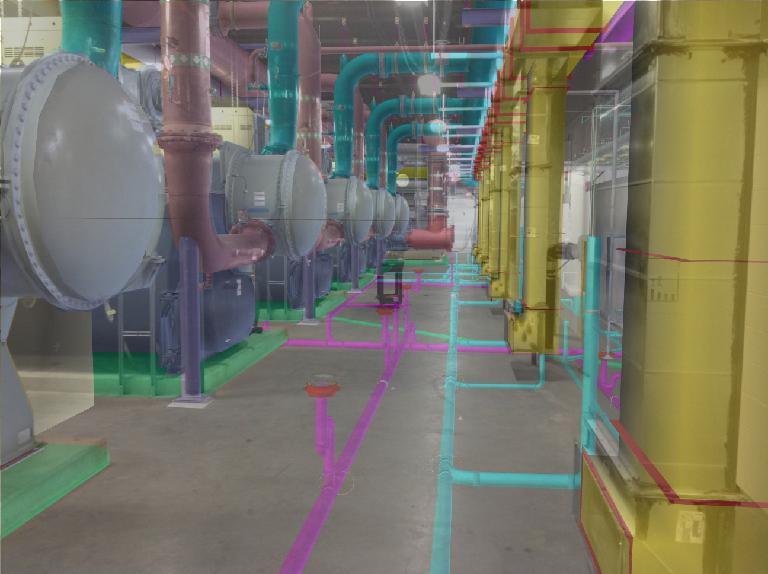}
\includegraphics[width=.325\linewidth,clip=true, trim=0 100 0 0]{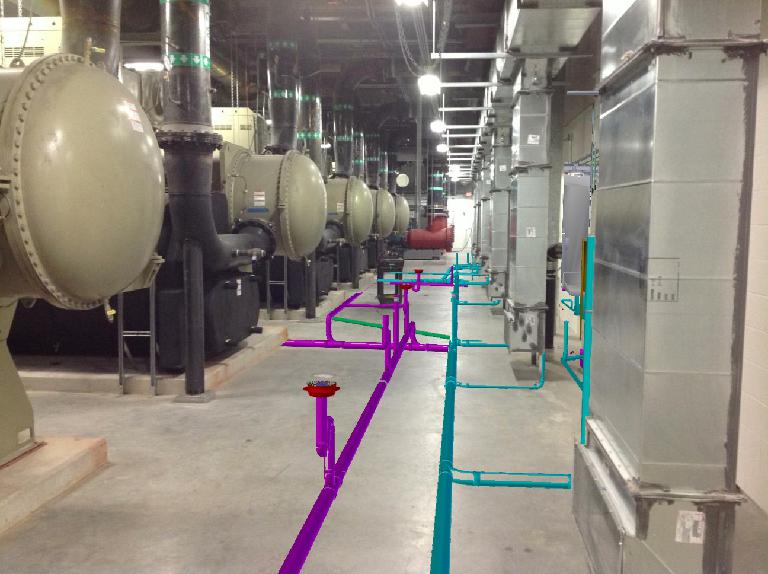}\\
\includegraphics[width=.325\linewidth,clip=true, trim=0 100 0 0]{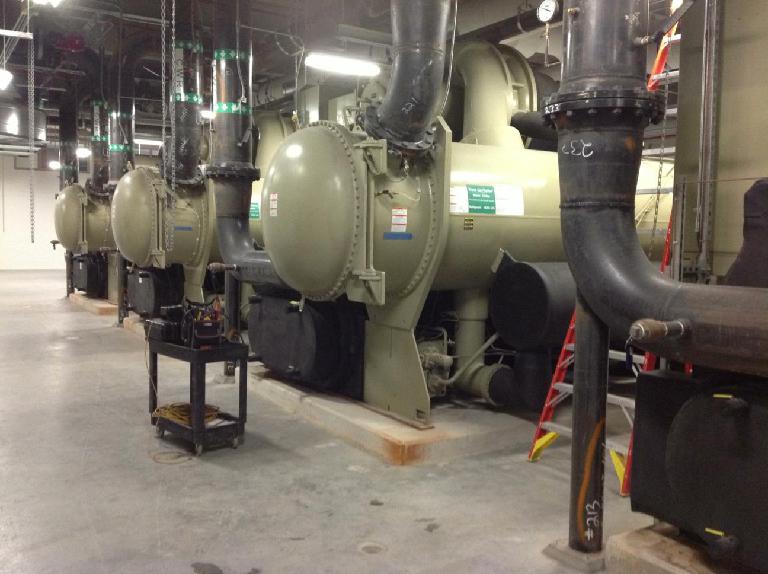}
\includegraphics[width=.325\linewidth,clip=true, trim=0 100 0 0]{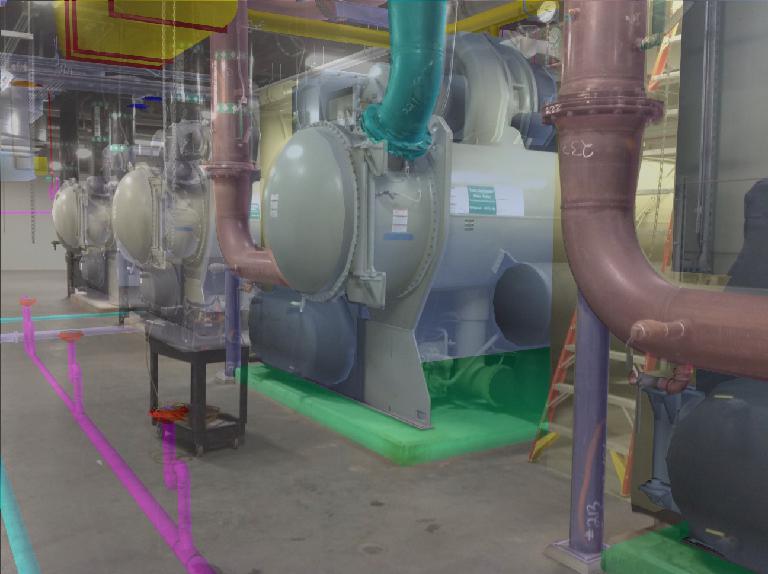}
\includegraphics[width=.325\linewidth,clip=true, trim=0 100 0 0]{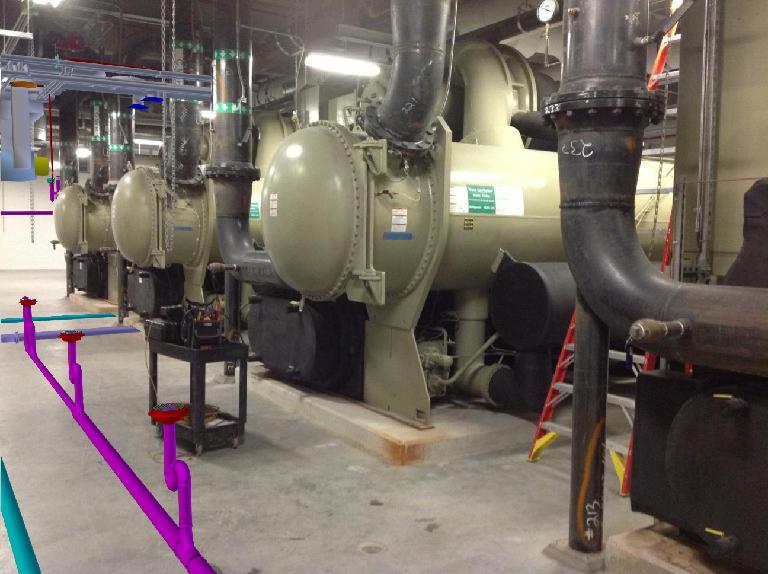}
\caption{Our system is applicable indoors and can register images with textureless surfaces (top row), repetitive structures, and significant occlusion (middle/bottom rows). Hidden or highly cluttered building systems, such as mechanical, electrical, and plumbing components can be visualized and annotated. Components can be filtered based on visibility and type (middle and bottom right).}
\label{fig:indoor}
\end{figure}
%-------------------

Subjects also completed an exit Likert questionnaire about the features of ConstructAide (e.g. ``Rate the usefulness of the [\ \ \ ] tool (1=poor, 5=excellent)''). Responses are summarized in Fig~\ref{fig:studyresponses}. Consistent with interview responses, subject found the 4D navigation feature to be the most useful and the construction error assessment tools to be the least useful. Occlusion removal and progress monitoring features were generally well-liked. Finally, all users responded ``yes'' to the exit question ``would you use this software for your field of work?''

%--------------------------------------
\newtext{
\subsection{Applications}

Beyond the primary use of construction monitoring, we also demonstrate other practical use-cases for ConstructAide:

\boldhead{Architectural visualizations for homeowners and clients} Pre-visualization of interior and exterior finishes, colors, or styles. Our system allows buyers take images from their desired viewpoints and obtain photorealistic renderings of different alternatives on their desired building components (e.g. choosing different tiles/ceramics and wallpapers for a bathroom). Our system eliminates the need for the physical mockups or reviewing samples, and instead allows the end users to review alternatives from one or many viewpoints of their choice (see Figure \ref{fig:timelapse}).

\boldhead{Performance, liability and dispute resolution} Visual records of a project's lifecycle can be extremely useful for handling legal and contractual disputes, but analyzing and jointly visualizing the photographs with project 3D models has previously been difficult. ConstructAide provides new solutions in this space, as echoed by subjects during our pilot interview.

\boldhead{Facility management} Despite efforts to link product data and maintenance schedules to BIM, these models are rarely used for operation and maintenance purposes. One issue is that today's smartphones do not have the capacity to directly provide access to BIM when/where such information is needed. The application of BIM with mobile augmented reality has been limited to proof-of-concepts due to challenges of using location tracking systems indoors.  Our solution can be used alongside current practices by allowing users take pictures and immediately analyze and visualize the site on their smart devices. To minimize scene clutter in presence of a large number of building elements in a model, our tools can filter visualizations based on element type or the task in hand. We show this method works both for indoors (Fig.~\ref{fig:indoor}) as well as outdoor scenes.

}

\section{Model-assisted Structure from Motion}
%-------------------------------------------------------------------------------
\label{sec:MeshSFM}

The availability of inexpensive and high-resolution mobile devices equipped with cameras, in addition to the Internet has enabled contractors, architects, and owners the ability to capture and share hundreds of photos on their construction sites on a daily basis. These site images are plagued with problems that can be difficult for existing SfM techniques, such as large baselines, moving objects (workers, equipment, etc.), and the constantly changing geometry/appearance of a construction site. \newnewtext{Furthermore, current SfM approaches are designed to work with hundreds of photographs of a static scene in which there is very high spatial density among the camera positions and view directions. Taking this many photos regularly of a construction site is not practical; a majority of site photos come from a few fixed-position, time-lapse cameras.} To overcome these issues, we propose a user-assisted SfM pipeline in which the user provides an accurate initial camera pose estimate (through mesh-image correspondences) which drives the remainder of the registration process.

\begin{figure}[t]
\includegraphics[width=\linewidth]{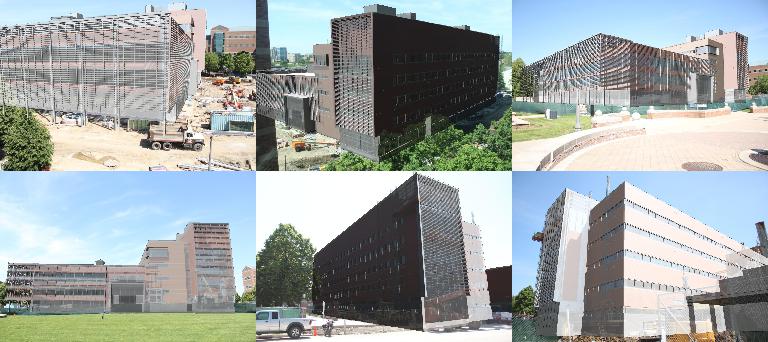}
\caption{Registrations estimated with Model-assisted SfM.}
\label{fig:registration}
\end{figure}

As in typical SfM algorithms, the result is a set of camera intrinsic and extrinsic parameter estimates as well as a sparse point cloud (triangulated feature correspondences). We are primarily interested in the camera parameters as this provides a registration of the 3D model to each photo. To a lesser extent, we use the point cloud when determining static occlusions (Sec~\ref{sec:4D:occ}). Fig~\ref{fig:registration} shows example registrations obtained using our method on a construction dataset.

Our goal is to find the proper camera parameters (intrinsic and extrinsic) as to register the virtual cameras with the {\it Euclidean} 3D model for each image. Here, we use the term {\it Euclidean} to represent the similarity transformation that maps an image-based 3D model into a measurable coordinate system for Engineering applications. We model intrinsic parameters using a three parameter pinhole model with variable focal length and two radial distortion coefficients, and assume that the principal point is in the center of the image and that pixels are square.

%-------------------
%-------------------

To begin the registration process, a user chooses one image from the collection (denoted throughout as an {\it anchor camera}) and selects 2D locations in the image and corresponding 3D points on the mesh model\footnote{Using the semantic/timing information present in the building information model, our interface allows for future building components to be hidden so that model can be easily made to match a photo taken at any point during the construction progress}. Our interface facilitates this selection by allowing the users to quickly navigate around the mesh. Given at least four corresponding points, we solve for the six-parameter extrinsic parameters of the camera -- three rotation and three translation parameters -- that minimize reprojection error using Levenberg-Marquardt (also called the Perspective-$n$-Point, or P$n$P problem). During this optimization, we fix the intrinsic parameters to have no radial distortion, and the focal length is obtained either from EXIF data or initialized such that the field of view is 50${}^\circ$. Prior to optimization, the camera parameters are initialized using the pose of of the model in the GUI (see supplemental video).

Choosing the anchor camera is important: this image should contain a sufficient view of the mesh such that many corresponding points in the image and on the model are clearly visible. This is typically straightforward as construction imagery focuses on a particular object (e.g. a building) and empirically we find many user-selected points to be visible in nearly all frames.

Knowledge of mesh-to-photo registration for {\it one} image doesn't help the SfM process as much as one might expect, but it does eliminate the coordinate system ambiguity (gauge transformation), and we later show how the 2D-to-3D correspondences can constrain and improve SfM estimates (see Tables~\ref{tab:quant_construction} and ~\ref{tab:middlebury_mvs}). Photos of construction sites are typically object/building-centric, so many cameras will be viewing the same object (many SfM methods cannot make this assumption, i.e. when reconstructing Rome~\cite{buildingRome}). Thus, the anchor camera can constrain many other images in the collection.

%Because the user has given us input about this registration, we can be reasonably certain about the accuracy of the camera pose for this image. We call these user-guided registrations {\it anchor} cameras, And we leverage this assumption to provide additional constraints for both registering images with wide baselines as well as the bundle adjustment.

%-------------------
\begin{algorithm}[t]
 \KwIn{3D model, set of unregistered images $\mathcal{U}$}
 \KwOut{set of registered images $\mathcal{R}$}
 $\mathcal{R} \gets \varnothing$\\
 Choose initial image $I$ from $\mathcal{U}$\\
 Select 2D-3D point correspondences between I and 3D model\\
 Solve for camera parameters\\
 $\mathcal{R}\gets \mathcal{R} \cup I$, \ $\mathcal{U}\gets \mathcal{U} \setminus I$\\
 Compute feature matches for each image pair in collection\\
 \While{$\mathcal{U} \neq \varnothing$ }{
  \ForEach{$R \in \mathcal{R}, $U$ \in \mathcal{U}$}{
    \If{Homography between $R$ and $U$ fits well}{
      Transfer selected 2D correspondences from $R$ to $U$
      Solve for $U$'s camera, $\mathcal{R}\gets \mathcal{R} \cup U$, \ $\mathcal{U}\gets \mathcal{U} \setminus U$
    }
  }
  Perform constrained bundle adjustment on all images in $\mathcal{R}$
  Identify $U'\in\mathcal{U}$ with most triangulated features (tracks)\\
  \eIf{$U'$ has enough matched tracks}{
   Triangulate tracks in $\mathcal{R}$ corresponding to features in $U'$
   }{
   Identify $U'\in\mathcal{U}$ with fewest matched tracks\\
   Select correspondences between $U'$ and 3D model\\
  }
  Solve for $U'$ camera, $\mathcal{R}\gets \mathcal{R} \cup U'$, \ $\mathcal{U}\gets \mathcal{U} \setminus U'$\\
  Perform constrained bundle adjustment on all images in $\mathcal{R}$
 }
\caption{Model-assisted SfM}
\label{alg:overview}
\end{algorithm}
%-------------------

%-------------------
\begin{table*}[t]
{\small
\begin{center}
\begin{tabular}{|c c|| c|c|c|c || c|c|c|c || c|c|c|c |} \hline
& & \multicolumn{4}{c||}{Rotational error (degrees)} & \multicolumn{4}{c||}{Translational error (meters)} & \multicolumn{4}{c|}{Reprojection error (\% width)} \\
Dataset & \# images & Ours & VSfM  & PS & PS-ICP & Ours & VSfM & PS & PS-ICP & Ours & VSfM & PS & PS-ICP \\ \hline \hline
Northwest A & 15 & \hili{0.67} & 2.28 & 8.79 & 79.40 & \hili{1.91} & 2.51 & 6.99 & 10.19 & \hili{9.34} & 22.70 & 23.26 & 52.96\\ \hline
Northwest B & 160 & 0.36 & \hili{0.30} & 0.31 & 5.13 & \hili{0.22} & 0.24 & 0.24 & 2.43 & 1.41 & \hili{0.87} & 0.94 & 17.93\\ \hline
West & 26 & \hili{1.20} & 1.81 & 1.67 & 20.02 & \hili{0.21} & 0.53 & 1.16 & 1.97 & \hili{1.37} & 1.96 & 3.32 & 20.25\\ \hline
Northeast & 22 & \hili{0.17} & 1.23 & 1.21 & 6.22 & \hili{0.15} & 1.34 & 1.14 & 9.08 & \hili{0.50} & 3.54 & 3.12 & 17.65\\ \hline
Basement & 10 & \hili{1.70} & 137.90 & 12.45 & 3.29 & \hili{0.44} & 8.15 & 1.56 & 1.55 & \hili{1.53} & 45.61 & 8.22 & 9.67\\ \hline
Southeast & 25 & \hili{0.31} & 0.73 & 0.94 & 5.00 & \hili{0.07} & 0.72 & 1.97 & 2.16 & \hili{0.52} & 1.86 & 3.63 & 9.49\\ \hline
\end{tabular}
\end{center}
}
\caption{Comparison of our method against existing approaches using real-world construction data on an instructional facility.}
\label{tab:quant_construction}
\end{table*}
%-------------------

After the first image is registered with the mesh model, we proceed by iteratively registering other images in the collection. Our approach is similar to many existing structure-from-motion algorithms, but with several important differences that leverage the anchor camera. Following existing structure-from-motion methods~\cite{SNAVELY-IJCV08}, we detect and match interest points across all images in the collection, and prune the matches by estimating the Fundamental matrix between image pairs using RANSAC. Different from existing methods, we then search for images which match the anchor image well up to a single homography (80\% of matched features are required as inliers), warp the selected 2D points from the anchor image to these images, and solve the P$n$P problem for each of these images using the known 3D correspondences to register nearby images (excluding points that fall outside the image; if fewer than four remain, we do not register the image). This is particularly useful for construction images, as many can be taken from roughly the same viewpoint with only focal length and rotational differences, such as those from a mounted camera.

Among all of the registered images, we perform one round of {\it constrained bundle adjustment}. As in most reconstruction approaches, our bundle adjustment optimizes over extrinsic/intrinsic camera parameters and triangulated 3D points; however, points triangulated using the anchor camera are constrained to lie along the anchor camera's ray, and we do not adjust the pose of the anchor camera (but intrinsics may change); refer to the supplemental document for details. We do not triangulate matched features corresponding to rays less than two degrees apart to avoid issues of noise and numerical stability. If no matches are triangulated, bundle adjustment is skipped.

One point of consideration is that the content of construction images will change drastically over time, even from similar viewpoints. However, many elements in these photos remain constant (ground/road, background trees/buildings) and we have found -- based on our datasets -- that features on these objects seem generally sufficient drive our SfM over long timescales. However, if automatic registration
fails, new anchor image(s) can easily be added.

%-------------------
\begin{table*}
{\small
\begin{center}
\begin{tabular}{|c c|| c|c|c || c|c|c || c|c|c |} \hline
& & \multicolumn{3}{c||}{Rotational error (degrees)} & \multicolumn{3}{c||}{Translational error (unitless)} & \multicolumn{3}{c|}{Reprojection error (\% width)} \\
Dataset & \# images & Ours &  VSfM  & PS & Ours & VSfM & PS & Ours & VSfM & PS\\ \hline \hline
temple (medium) & 47 & \hili{2.03} & 2.47 & 2.84 & 0.05 & \hili{0.02} & 0.04 & \hili{1.94} & 2.32 & 2.02\\ \hline
temple (small) & 16 & \hili{2.34} & 2.72 & 3.35 & \hili{0.05} & 0.11 & 0.07 & 2.01 & \hili{1.91} & 2.13\\ \hline
dino (medium) & 48 & 2.43 & \hili{0.79} & 6.27 & \hili{0.06} & 0.09 & 0.33 & 0.75 & \hili{0.72} & 1.78\\ \hline
dino (small) & 16 & 3.06 & 6.46 & \hili{0.76} & \hili{0.08} & 0.50 & 0.30 & \hili{1.04} & 4.01 & 1.13\\ \hline
\end{tabular}
\end{center}
}
\vspace{-1mm}
\caption{Evaluation using ground truth data from the Middlebury Multi-View Stereo dataset.}
\label{tab:middlebury_mvs}
\end{table*}
%-------------------

Next, we search for other images with a sufficient number of features corresponding to existing tracks, i.e. matched features common to two or more registered images; such features can be triangulated. We choose the image that has the fewest matches over a threshold (60 in our implementation) to ensure a good match and potentially wide baseline. This camera is registered by solving a constrained P$n$P problem using its 2D matches corresponding to the triangulated 3D tracks, made robust with RANSAC (inliers considered within 1\% of the image width). We also use the anchor camera to improve the P$n$P solution: using the Fundamental matrix between the anchor camera image and the image that is currently being registered, epipolar lines are computed corresponding to the user-selected 2D locations in the anchor image; the corresponding 3D mesh locations are then constrained to lie nearby these lines (based on reprojection error). Given a set of $k$ 3D points $X=\{X_1, \ldots, X_k\}$ and their corresponding projected pixel locations $u=\{u_1, \ldots, u_k\}$ and epipolar lines $e=\{e_1, \ldots, e_k\}$, we search for a 3D rotation ($R$) and translation ($t$) that jointly minimizes reprojection error as well as the point-to-line distance from projected points to their corresponding epipolar lines:
\begin{align}
\argmin_{R,t} \sum_i || x_i - u_i || + pld(x_i, e_i), \nonumber \\
\text{where: } x_i = \text{project}(R X_i + t, f)
\label{eq:contrainedPnP}
\end{align}
where $\text{project}(X,f)$ projects 3D locations into the plane according to focal length $f$, and $pld(x,l)$ computes the shortest distance from pixel location $x$ to the line specified by $l$. In our experience, this strategy helps avoid errors due to noisy camera estimates and triangulations.

In the case that not enough features in unregistered images match existing tracks in registered images, we choose the image with the {\it least} matched track of feature points. The user then specifies 2D locations in this image corresponding to 3D mesh locations selected in the anchor image\footnote{This is purely an image-based task, as the 3D positions do not need to be specified again.}, and this image is registered again by solving P$n$P. This happens typically if the image graph, or sets of tracks through the image collection, is disjoint. The image with the least matched tracks is chosen with the goal of connecting the graph, or at the very least, adding an image with large baseline. Since the user assisted in registering the chosen image, this camera is also designated as an anchor camera. After this camera is registered, another round of constrained bundle adjustment is performed. Until all images have been registered, this process is repeated. See Algorithm~\ref{alg:overview} for an overview.

%-------------------------------------------------------------------------------
\subsection{Experiments}
%-------------------------------------------------------------------------------

We hypothesize that knowing at least one camera's pose (as in our method) should aid camera pose and reconstruction estimates, as compared to blind, automatic SfM techniques. To test our hypothesis (and accuracy of registration), we compared our estimates to ground truth camera poses as well as camera pose estimates from established SfM methods. In total, we tested 10 different photo collections falling into two categories: real-world construction site images and object-centric images from the Middlebury Multiview Stereo dataset~\cite{middlebury_mvs}. We chose this data for several reasons: construction site data allows us to quantify error on real-world sites, the data vary widely in appearance, baseline, and number of photos, testing the limits of our method, and we require a corresponding mesh-model (available for our construction data, and obtainable for the Middlebury data). We compare our method to Wu's VisualSfM~\shortcite{vsfm1,vsfm2} and Photosynth\footnote{\url{http://photosynth.net}. Camera information extracted using the Photosynth Toolkit: \url{https://code.google.com/p/visual-experiments/}}. While both methods are based on the method of Snavely et al.~\cite{phototourism}, we found the estimates to be quite different in some cases most likely due to differences in implementation (e.g. Photosynth uses a different feature matching scheme than VisualSFM\footnote{\url{http://en.wikipedia.org/wiki/Photosynth}}).

\boldhead{Construction Site Evaluation}
We first test our method on real-world construction data. Ground truth camera pose estimates do not exist for this data, so we create ground truth data by manually calibrating five of the images in each dataset (images are chosen for dataset coverage). Corresponding 2D and 3D locations are chosen by hand, allowing us to solve for the true camera pose. At least four pairs of corresponding points must be chosen, and the set of chosen points should not be co-planar to avoid projective ambiguity. As our method requires the same ground truth calibration for at least one of the images (during initialization), we ensure that the images calibrated in our method are not used when creating ground truth (and thus not compared to).

\newtext{In order to test the limits of each of the evaluated algorithms, we ensure that our datasets vary significantly in spatial and temporal sampling of photos. Each dataset required between two and four anchor images depending on the spatial density.} For each photo collection, we process the images with our model-assisted SfM technique (Sec~\ref{sec:MeshSFM}) as well as VisualSfM and Photosynth (denoted as {\bf VSfM} and {\bf PS} onward). Since the models produced by VSfM and PS are not in the same coordinate system as the ground truth data, we align them with a simple procedure: (a) triangulate a set of points (hand-selected for accuracy) using both the ground truth cameras and VSfM's cameras, (b) find the similarity transformation (scale, rotation, translation) that minimizes the squared distance between the point sets, and (c) apply this transformation to VSfM's cameras. The same procedure is applied to the result from PS. For nearly all datasets, the mean squared error is $<0.01$m, ensuring a good fit. There is no need to adjust the pose estimates from our method as our estimates are already in the 3D model's coordinate system.

For additional comparison, we also match the coordinate system of PS results to the ground truth by matching all triangulated features with points sampled from the 3D model using the iterative closest point algorithm; we call this method {\bf PS-ICP}.

Between each of the methods and the ground truth cameras, we compute three error measures: rotational difference (angle between viewing directions), translational difference (distance between camera centers, in meters), and reprojection error of seven hand-selected ground truth 3D locations (selected for wide coverage of the model). Table~\ref{tab:quant_construction} shows the results of this experiment on the six construction site photo collections. The errors shown are averaged over all five of the ground truth calibrations.
%We also show several qualitative registration results in Fig~\ref{fig:qual_construction}.

\boldhead{Middlebury Evaluation}
We also test our method and others against ground truth camera pose from the Middlebury Multiview Stereo datasets. We investigate four of the datasets (dino and temple datasets, the medium and small collections), and compare our method with VisualSfM (VSfM) and Photosynth (PS). As in the construction data experiment, we compute rotational, translational, and reprojection error. Since we now have ground truth data for each of the images in the dataset, we compute the average error over all images in the dataset (excluding any that were not successfully registered by a particular algorithm). Table~\ref{tab:middlebury_mvs} shows the results.

\boldhead{Discussion}
In both experiments, we observe that our model-assisted SfM technique typically outperforms existing methods in the three error measures (although the compared methods do not have access to a 3D model during registration). Incorporating 3D model data into the SfM process can be beneficial at a low cost to the user, even if the model is incomplete/inexact. We see that the results are fairly consistent across the two experiments, indicating that our method might be suitable for ``object-sized'' data as well.

These experiments suggest that our method may perform better than other techniques for smaller image collections with wider baselines. For example, the Basement sequence contains only ten photos from significantly different viewpoints. Our method is able to achieve low error while other techniques cannot handle this sparse sampling of photographs. Thus, our system is uniquely suited for a small number of viewpoints with a dense time sampling (e.g. a few time-lapse cameras around a construction site) -- annotation, rendering, occlusion detection, and temporal navigation are all still possible given these kinds of datasets. These scenarios are what our system is designed for, whereas existing SfM techniques require spatially dense images (also suggested quantitatively by our evaluation). The proposed interface and visualizations are more compelling with dense time data but possible otherwise.

\begin{figure}[h]
\includegraphics[width=\linewidth]{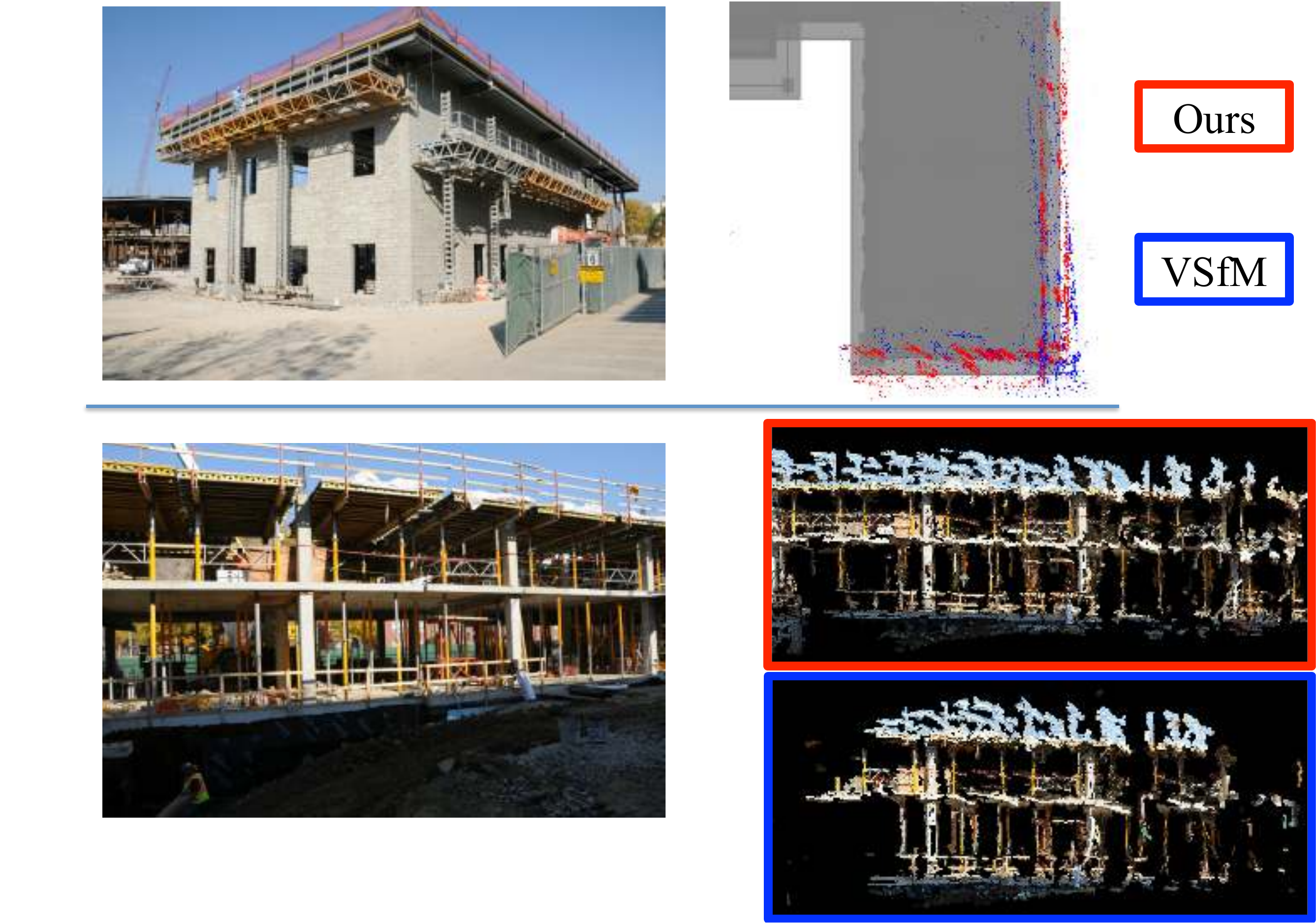}
\caption{Our SfM method typically produces more robust estimates than existing automatic approaches, resulting lower reconstruction error and less drift (top row; VSfM's point cloud propagates error away from the building corner) as well as more dense multi-view stereo reconstructions (bottom row; using the PMVS2 software of Furukawa and Ponce~\protect\shortcite{Furukawa:2010}).}
\label{fig:drift_completeness}
\end{figure}

For larger, more complete collections, existing automatic techniques methods may suffice, although a manual coordinate-system registration process must still be used to bring the cameras into the 3D model's coordinate system. Our system can also handle spatially dense sets (e.g. Table~\ref{tab:quant_construction} Northwest B) but makes no explicit requirements and typically yields improvements and reduced accumulated error compared to existing methods (Fig~\ref{fig:drift_completeness}).

%In Fig~\ref{fig:drift_completeness}, we also demonstrate that the improved camera pose estimates computed by our method can lead to better dense reconstructions, and helps to eliminate accumulated error (drift).

%-------------------
%-------------------

%-------------------------------------------------------------------------------
\section{Limitations}
%-------------------------------------------------------------------------------

Currently, our system relies on a small amount of user interaction to provide pixel-precise registration. Our method automates several tasks (such as the handling of occlusions and computing sun direction), but if errors occur, the user must correct these using the smart-selection tools provided by our interface.

\newtext{Our technique requires accurate and complete BIM data which are typically only available for commercial construction sites. To get the most out of our system, the BIMs must also contain semantic information such as materials, scheduling information, and sufficient levels of detail; the input photographs should be both spatially and temporally dense as well. As such, our system may not be suitable for certain photo collections and construction sites.} While our user study provided preliminary information validating our system's usefulness and applicability, a comprehensive study with more users is necessary to evaluate the user interface and practicality in the field.

\boldhead{Failure cases} As with many structure from motion techniques, our method may fail to register an image in the presence of either inadequate/inaccurate feature matches or a poor initialization. In these cases, we allow the user to fix the registration by selecting 2D point correspondences, as described in Sec~\ref{sec:MeshSFM}. Furthermore, occlusion estimation may fail due to too few nearby photographs. For example, our static occlusion detection can fail if a SfM point cloud is too sparse or unavailable, and our dynamic occlusion detection can fail if no nearby viewpoints/images exist in the photo collection.

%-------------------------------------------------------------------------------
\section{Conclusion}
%-------------------------------------------------------------------------------

We have demonstrated a system that aligns 4D architectural/construction models and photographs with high accuracy and minimal user interaction. Our system is quick and easy to use, and enables many valuable job-site visualization techniques. Our interface can be used to navigate the construction site both forwards and backwards in time, assess construction progress, analyze deviations from the building plans, and create photorealistic architectural visualizations  without the time and training required to learn complicated CAD, modeling and rendering software. Our proposed model-assisted SfM algorithm and outperforms existing techniques achieves precise registration, enabling semantic selection tools and accurate visualization. Using this data, we show that occlusions can be reasoned accurately, and materials and lighting can be extracted from the plan data for rendering.

\newtext{Many technical challenges remain for further automation. For example, assessing construction progress and errors automatically would be very useful for site inspectors. Automatically registering a model to a photo (similar to~\cite{Aubry13}) is another interesting avenue for future work, especially in the presence of incomplete or inaccurate building models. Optimizing these processes (such as SfM) is also necessary for enabling our system on mobile devices so that it can be used in real-time on job sites.
}

%Future directions of work include further automation, e.g. automatically registering a model to a photo (similar to~\cite{Aubry13}), automatically detecting construction errors or assessing progress, and new ways to leverage 3D models during the SfM process. Our system could also be used to visualize archeological and historical sites through archival photos and/or paintings.

%We will also look into enabling mobile augmented reality on the job site without the need for keeping the very large files associated with the most updated 4D architectural/construction models or rendering these files on the device.
%Using our system, project managers and facility owners can visualize and assess jobs with unprecedented ease.

%-------------------------------------------------------------------------------
\section*{Acknowledgements}
%-------------------------------------------------------------------------------
This work was funded in part by NSF award CMMI-1360562 and the NSF GRFP. We thank University of Illinois Facilities, Turner Construction, and Zachry Construction for providing access to data, as well as Siddharth Kothari for help with data curation.

%-------------------------------------------------------------------------------
{
%\renewcommand{\baselinestretch}{0.85}
%\small
\bibliographystyle{acmsiggraph}
\bibliography{SfM,aec,sg14refs}
}%-------------------------------------------------------------------------------

\end{document}